%% file: JpsiMult_pp13.tex
\documentclass[ALICE,manyauthors]{cernphprep}

\usepackage[comma,square,numbers,sort&compress]{natbib}
\usepackage{hyperref}
\usepackage{color}
\usepackage{lineno}

\usepackage[T1]{fontenc}
\usepackage{orcidlink}

\DeclareGraphicsExtensions{.pdf} 
\graphicspath{{./figs/}}

\usepackage[T1]{fontenc}

\newcommand{\sqrts}[1]   {\mbox{$\sqrt{s} = #1$ TeV}\xspace}
\newcommand{\sqrtsnn}[1] {\mbox{$\sqrt{s_{\mathrm{NN}}} = #1$ TeV}\xspace}

\newcommand{\Jpsi}{{\ensuremath{\boldsymbol{\mathrm{J}/\psi}}}\xspace}
\newcommand{\diele}{\ensuremath{{\mathrm e}^{+}{\mathrm e}^{-}}\xspace}

\newcommand{ \mathrms }[1]{r.m.s.}

\input{commands}

\begin{document}%

\begin{titlepage}
\PHyear{2026} 
\PHnumber{109}      
\PHdate{03 April}  
%

\title{Multiplicity dependence of prompt and non-prompt $\boldsymbol{\mathrm{J}/\psi}$ production at midrapidity in pp collisions at $\mathbf{\sqrt{\textit{s}} = 13}$ TeV}
\ShortTitle{Multiplicity-dependent prompt and non-prompt \jpsi production at \sqrts{13}}   

\Collaboration{ALICE Collaboration\thanks{See Appendix~\ref{app:collab} for the list of collaboration members}}
\ShortAuthor{ALICE Collaboration} 

\begin{abstract}
The yields of prompt and non-prompt \jpsi and the fraction of non-prompt \jpsi are measured at midrapidity ($|y|<0.9$) via the dielectron decay channel as a function of the midrapidity charged-particle multiplicity ($|\eta|<0.9$) in pp collisions at $\sqrt{s} = 13$ TeV. The \jpsi yields and the multiplicity are normalized by their average value in inelastic collisions. The multiplicity-dependent yield ratio between prompt \jpsi and D$^0$ is reported. The multiplicity is further divided into three azimuthal regions with respect to the \jpsi momentum: toward the \jpsi emission direction, transverse, or opposite to it. A stronger-than-linear increase of the self-normalized yields is observed for both prompt and non-prompt \jpsi production, with similar trends. This behaviour is also observed in the toward region, while a weaker increase is observed in the transverse and away regions. 
\end{abstract}

\end{titlepage}
\setcounter{page}{2}

%
%
\section{Introduction}
\label{sec:intro}
\input{./A_Introduction.tex}

\section{Experimental setup and data samples}
\label{sec:data}
\input{./B_Experiment.tex}

\section{Analysis}
\label{sec:ana}
\input{./C_Analysis.tex}

\section{Results and discussion}
\label{sec:meas}
\input{./D_Results.tex}

\section{Summary and conclusions}
\label{sec:conc}
\input{./E_Summary.tex}

%
%

\newenvironment{acknowledgement}{\relax}{\relax}
\begin{acknowledgement}
\section*{Acknowledgements}

We are grateful to E. Levin, M. Siddikov, R. Venugopalan, K. Watanabe, J. Zhao, and K. Werner for sending us the predictions of their models and for clarifications in this regard. 
%
\input{fa_2026-02-26_Opt_C.tex}    
\end{acknowledgement}

\bibliographystyle{utphys}   
\bibliography{Jpsi}

\newpage
\appendix
\section{Discussion on the inclusion of the decay daughters in the multiplicity}
\label{sec:daughters}
\input{./AppendixA_DaughtersRemoval.tex}

\section{Additional PYTHIA curves}
\label{sec:PythiaAdditional}

\input{./AppendixB_AdditionalModels.tex}

\clearpage

\section{The ALICE Collaboration} 
\label{app:collab}
\input{Alice_Authorlist_2026-02-26_Opt_C.tex}  

\end{document}

%% file: commands.tex
\newcommand{\pt}           {\ensuremath{p_{\rm T}}\xspace}

\newcommand{\mee}          {\ensuremath{m_{\rm e^+e^-}}\xspace}

\newcommand{\onethreeH}       {$\sqrt{s_{\mathrm{NN}}}~=~130$~Ge\kern-.1emV\xspace}
\newcommand{\nineH}        {$\sqrt{s}~=~0.9$~Te\kern-.1emV\xspace}
\newcommand{\oneninesix}  {$\sqrt{s}~=~1.96$~Te\kern-.1emV\xspace}
\newcommand{\twoH}         {$\sqrt{s}~=~200$~Ge\kern-.1emV\xspace}
\newcommand{\twosevensix}  {$\sqrt{s}~=~2.76$~Te\kern-.1emV\xspace}
\newcommand{\five}         {$\sqrt{s}~=~5.02$~Te\kern-.1emV\xspace}
\newcommand{\fivefive}         {$\sqrt{s_{\mathrm{NN}}}~=~5.5$~Te\kern-.1emV\xspace}
\newcommand{\seven}        {$\sqrt{s}~=~7$~Te\kern-.1emV\xspace}
\newcommand{\eight}        {$\sqrt{s}~=~8$~Te\kern-.1emV\xspace}
\newcommand{\thirteen}     {$\sqrt{s}~=~13$~Te\kern-.1emV\xspace}
\newcommand{\thirteensix}     {$\sqrt{s}~=~13.6$~Te\kern-.1emV\xspace}
\newcommand{\fourteen}     {$\sqrt{s}~=~14$~Te\kern-.1emV\xspace}
\newcommand{\twosevensixnn}{$\sqrt{s_{\mathrm{NN}}}~=~2.76$~Te\kern-.1emV\xspace}
\newcommand{\fivenn}       {$\sqrt{s_{\mathrm{NN}}}~=~5.02$~Te\kern-.1emV\xspace}

\newcommand{\GeVc}         {Ge\kern-.1emV/$c$\xspace}
\newcommand{\MeVc}         {Me\kern-.1emV/$c$\xspace}
\newcommand{\TeV}          {Te\kern-.1emV\xspace}
\newcommand{\GeV}          {Ge\kern-.1emV\xspace}
\newcommand{\keV}          {ke\kern-.1emV\xspace}
\newcommand{\MeV}          {Me\kern-.1emV\xspace}
\newcommand{\GeVmass}      {Ge\kern-.2emV/$c^2$\xspace}
\newcommand{\MeVmass}      {Me\kern-.2emV/$c^2$\xspace}

\newcommand{\jpsi}         {\ensuremath{\text{J}/\psi}\xspace}


%% file: A_Introduction.tex

Charmonia, such as \jpsi mesons, are bound states of charm and anticharm quarks. They can be produced directly in the collision, arise from the decay of a higher charmonia state, or from the weak decays of open-beauty hadrons. In the first two cases, the charmonia are called prompt, and they are called non-prompt otherwise. While perturbative quantum chromodynamics (pQCD) calculations can describe the production cross section of a heavy-quark pair, its hadronization into a prompt \jpsi meson cannot be computed from first principles and has to be described by phenomenological models. One of them is the Improved Color Evaporation Model (ICEM)~\cite{Ma:2016exq}, where the transition  of a produced heavy-quark pair to a colorless state, by emitting soft gluons, happens with a fixed probability. Alternatively, in the Non-Relativistic QCD (NRQCD) model~\cite{Bodwin:1994jh, Ma:2014mri}, the heavy-quark pair may be produced in either a color-singlet or a color-octet state. The transition from these states to the physical quarkonium is described by Long-Distance Matrix Elements, which are extracted from global fits to data~\cite{Butenschoen:2010rq}. These calculations have also been complemented by taking into account the transverse momentum distribution of the initial gluons in the $k_{\rm T}$-factorization scheme~\cite{Cheung:2018tvq,Baranov:2019lhm}, or by calculating the gluon momentum distributions within the Color Glass Condensate (CGC) framework~\cite{Ma:2014mri}. In the latter, also contributions from triple gluon fusion have been included~\cite{Motyka:2015kta,Levin:2018qxa}.

The hadronization of the b quark to produce a non-prompt \jpsi meson can be modelled by different mechanisms, such as fragmentation or coalescence. Recent measurements of baryon-to-meson ratios in the charm and beauty sectors in proton-proton (pp) collisions at \sqrts{13} by ALICE~\cite{ALICE:2023sgl, ALICE:2023wbx} and LHCb~\cite{LHCb:2023wbo} show a non-universality of these hadronization mechanisms compared to e$^+$e$^-$ and ep collisions, as well as a dependence on the multiplicity~\cite{ALICE:2021npz, LHCb:2022syj, LHCb:2023wbo}.

Investigating the dependence of \jpsi production on the charged-particle multiplicity (i.e. the number of charged particles produced in the collision) can provide valuable insights. Firstly, as most charged particles are produced in soft scatterings, the interplay between hard and soft particle production mechanisms can be investigated. Secondly, measurements of charged hadrons in high-multiplicity pp events at the LHC have shown signs of collectivity similar to those observed in the quark--gluon plasma (QGP)~\cite{Grosse-Oetringhaus:2024bwr}. The strange particle yields are found to continuously increase as a function of the multiplicity from pp to p–-Pb and Pb--Pb collisions~\cite{ALICE:2016fzo,ALICE:2022wpn,ALICE:2025cqy,ALICE:2025aqz}.
This motivates further studies of high-multiplicity events, also for events containing heavy-flavoured particles. In addition, the comparison between the multiplicity dependence of prompt and of non-prompt \jpsi production could help investigating the effects of the parton mass and of the hadronization mechanism on the measurements.

The multiplicity dependence of the self-normalized yields of \jpsi, $\psi(2S)$, and $\Upsilon(nS)$ mesons has been measured in pp collisions across various rapidity regions at the LHC by ALICE~\cite{ALICE:2012pet, ALICE:2020msa, ALICE:2021zkd, ALICE:2022yzs, ALICE:2022gpu, ALICE:2025fzz} and CMS~\cite{Chatrchyan:2013nza}, and at RHIC by PHENIX~\cite{PHENIX:2024dqs} and STAR~\cite{Adam:2018jmp, STAR:2025ywy}. The yields are self-normalized, i.e. they are normalized by their average values in inelastic events. A stronger-than-linear increase of the yields with the self-normalized multiplicity is observed when the quarkonium and the multiplicity are measured in the same rapidity region and the \jpsi decay daughters (two charged leptons) are included in the multiplicity  calculation~\cite{ALICE:2020msa, STAR:2025ywy,ALICE:2025fzz}. The increase is even stronger at higher quarkonium transverse momentum. However, when quarkonium and multiplicity are measured in different rapidity regions, the increase of the self-normalized yields with multiplicity is weaker. It is close to linear at LHC energies~\cite{ALICE:2022yzs, ALICE:2021zkd, ALICE:2022gpu}, and weaker than linear at RHIC energies~\cite{PHENIX:2024dqs}. A weaker-than-linear increase also occurs at RHIC energies when the quarkonium decay daughters are removed from the multiplicity calculation~\cite{PHENIX:2024dqs}. Several other measurements of hard particle production reported by ALICE, such as high-\pt hadrons~\cite{Acharya:2019mzb}, D-mesons~\cite{Adam:2015ota} and electrons from heavy-flavour decays~\cite{ALICE:2023xiu} in pp collisions at $\sqrt{s} = 5.02,7$ and 13 TeV also show similar stronger-than-linear increases with multiplicity. In contrast, electrons from W boson decays in pp collisions at $\sqrt{s} = 13$ TeV show a near-linear increase~\cite{ALICE:2026lnw}. 

The yield ratios of excited-to-ground quarkonium states have also been measured in pp collisions by ALICE~\cite{ALICE:2022gpu, ALICE:2022yzs}, LHCb~\cite{LHCb:2023xie, LHCb:2025xrb} and CMS~\cite{CMS:2020fae} at the LHC, and by PHENIX~\cite{PHENIX:2024dqs} and STAR~\cite{STAR:2025ywy} at RHIC. ALICE, PHENIX and STAR results show a trend compatible with a flat behaviour, while the more precise measurements by LHCb and CMS of the prompt charmonium and bottomonium yield ratios indicate a decrease of these ratios as a function of multiplicity. CMS has also measured in pp collisions at \sqrts{7}~\cite{CMS:2020fae} the $\Upsilon(nS)$ ratios as a function of the multiplicity in the azimuthal direction towards, transverse, or opposite to the $\Upsilon$ emission direction. No major differences were found between these three cases. The decrease was also found to be stronger in isotropic events compared to jet-like events, for which no modification of the ratio with multiplicity is visible, suggesting that the decrease depends on underlying event properties.

The multiplicity dependence of \jpsi and $\psi(2S)$ production was also measured in p--Pb collisions at $\sqrt{s_{\rm NN}} = 5.02$ and 8.16 TeV by ALICE~\cite{Adamova:2017uhu, Acharya:2020giw, ALICE:2022gpu}. At midrapidity, the increase in the self-normalized \jpsi yields as a function of the midrapidity multiplicity is also stronger than linear. The self-normalized \jpsi yield increases more weakly at forward rapidity in the proton-going direction than at backward rapidity in the Pb-going direction. The probed Bjorken-$x$ values of the partons probed in the Pb nucleus are lower in the former case. For both forward and backward rapidities, the increase of $\psi(2S)$ yield has been found compatible with the one from \jpsi within uncertainties.

Several theoretical models aim to describe the strong increase of the \jpsi yield with multiplicity. The influence of the multiple partonic interactions (MPI) on the multiplicity dependence is considered in these models~\cite{Ma:2018bax, Salazar:2021mpv,Siddikov:2019xvf, Gotsman:2020ubn,Terra:2023srs,Kopeliovich:2013yfa, Kopeliovich:2019phc,Ferreiro:2012fb,Bierlich:2022pfr, Christiansen:2015yqa,Werner:2023zvo, Werner:2023fne, Werner:2023jps, Werner:2023mod}. In addition, some models rely on initial-state effects inside the colliding protons, such as in the CGC framework. In this framework, the gluon density saturates due to non-linear effects at low $x$~\cite{Ma:2018bax, Salazar:2021mpv}, and the three-Pomeron fusion can have a strong influence at high multiplicity~\cite{Siddikov:2019xvf, Gotsman:2020ubn}. Other models rely on the Bjorken-$x$ dependent spatial distribution of gluons~\cite{Terra:2023srs}, or rely on parallels with the Glauber model in pA collisions~\cite{Kopeliovich:2013yfa, Kopeliovich:2019phc}. Other explanations include final-state effects, where the yield of soft particles could be additionally suppressed in high charged-particle density environments compared to low-density environments~\cite{Ferreiro:2012fb}. Event generators such as PYTHIA 8~\cite{Bierlich:2022pfr, Christiansen:2015yqa} or EPOS4~\cite{Werner:2023zvo, Werner:2023fne, Werner:2023jps, Werner:2023mod} typically implement several of the initial- and final-state effects in hard and soft MPI. These effects include, for example, reconnections between strings or hydrodynamic evolution in high charged-particle-density environments.

The ratio between prompt quarkonium and open charm production is a particularly appropriate discriminant because several experimental and theoretical uncertainties cancel out in the ratio. In Pb--Pb collisions, this ratio increases when going from semicentral to central collisions, as measured by ALICE at midrapidity at \sqrtsnn{5.02}~\cite{ALICE:2023gco, ALICE:2022wpn}, possibly due to regeneration of \jpsi mesons from uncorrelated c and $\rm \overline{c}$ quarks in a dense medium with a high number of $\rm c\overline{c}$ pairs produced in the collision. In LHCb fixed-target Pb--Ne collisions $\sqrt{s_{\rm NN}} = 68.5$ GeV, this ratio decreases with the system size~\cite{LHCb:2022qvj}. The multiplicity dependence of the ratio in pp collisions could provide information on whether dense medium effects could also be seen in small systems.

The hardest scattering in a collision typically results in outgoing high-\pt partons which fragment into jets, increasing the particle yield in a specific direction.
In contrast, the region azimuthally transverse to that direction probes mostly the soft MPI in the event, giving a cleaner estimate of the underlying event activity~\cite{Martin:2016igp,ALICE:2023csm,ALICE:2019mmy}. 
Therefore, the separation of the multiplicity estimators in azimuthal regions
can shed more light on the mechanisms causing the observed stronger-than-linear increase of the self-normalized \jpsi yield~\cite{Weber:2018ddv}. Particles associated with the \jpsi production can be found in a direction close to the one of the \jpsi meson. These include, for example, the particles produced inside a jet cone, or in the same decay process of a common mother particle such as a beauty hadron or higher mass charmonium. The multiplicity azimuthally transverse to the \jpsi emission direction mostly measures the underlying event activity. It might, however, contain initial-state or large-angle radiation produced in the same process as the one leading to the \jpsi production. Finally, the opposite region can include the recoil of the process leading to the \jpsi production. 
The correlation between the \jpsi production and the event activity has also been studied in pp collisions via the \jpsi-hadron azimuthal correlations~\cite{ALICE:2024yqd}, measured by ALICE  at \sqrts{13} and found in good agreement with PYTHIA, and the \pt fraction carried by the \jpsi meson in jets, measured by LHCb~\cite{LHCb:2017llq} and ALICE~\cite{ALICE:2026ywy} at \sqrts{13}, and CMS at \sqrts{5.02}~\cite{CMS:2021puf}. While some theoretical calculations can reproduce the latter measurement~\cite{Bain:2017wvk}, the jet fragmentation to the \jpsi meson is softer than predicted by PYTHIA.

This paper reports the measurement of prompt and non-prompt \jpsi production as a function of multiplicity, in pp collisions at a center-of-mass energy of $\sqrts{13}$. The data were collected with the ALICE detector during Run 2 of the LHC. The charged-particle multiplicity is measured in $|\eta|<0.9$, both in full azimuthal angle and separated into three azimuthal regions relative to the \jpsi emission direction. Section~\ref{sec:data} describes the experiment and the data sample used. Section~\ref{sec:ana} explains the analysis procedure. Section~\ref{sec:meas} presents the results and discusses their interpretation. Finally, a summary and conclusions are given in Section~\ref{sec:conc}.

%% file: B_Experiment.tex

The \jpsi mesons are reconstructed in the dielectron decay channel using the central barrel of ALICE~\cite{Abelev:2008aa, Abelev:2014ffa}. The detectors used for the track reconstruction are the Inner Tracking System (ITS)~\cite{Aamodt:2010aa} and the Time Projection Chamber (TPC)~\cite{Alme:2010ke}. They both have a full azimuthal coverage and are placed within a magnetic field of 0.5 T in the beam direction. The ITS is the detector closest to the interaction point. Due to its high spatial resolution, it plays a main role in the reconstruction of both the primary collision vertex and the secondary vertices from long-lived weakly decaying particles. It consists of two layers of Silicon Pixel Detectors (SPD), located at radial distances of 3.9 and 7.6 cm from the interaction point, respectively, followed by two layers each of Silicon Drift Detectors (SDD) and Silicon Strip Detectors (SSD). The TPC is the main ALICE tracking detector and also performs particle identification using the specific energy loss in its gas volume. It provides accurate momentum resolution and particle separation over a broad momentum range around midrapidity. 

The data samples used in this analysis have been recorded during the LHC Run 2 in pp collisions at a center-of-mass energy of $\sqrts{13}$ using several event triggers. A Minimum-Bias (MB) trigger requires the simultaneous detection of a signal in the V0A ($2.8 <\eta<5.1$) and V0C ($-3.7<\eta<-1.7$), two scintillator arrays with fast readout which together form the V0 detector~\cite{Abbas:2013taa}. A High-Multiplicity (HM) trigger requires the sum of the signals in the V0A and V0C (denoted as V0M) to be higher than a threshold value. It selects the 0.1\% of the collisions with the highest V0M signal. Therefore, the corresponding event class is called V0M~0--0.1\%.
Additionally, a trigger based on the Transition Radiation Detector (TRD)~\cite{ALICE:2017ymw} is also used. The TRD is made of 522 chambers distributed in the full azimuth within $|\eta|<0.84$. Each chamber contains a radiator, inducing transition radiation when ultra-relativistic electrons and positrons traverse it, as well as a drift chamber, where the energy deposition of the particles is measured. The TRD trigger is configured to select events containing at least one track with an online-computed \pt higher than 2~\GeVc and with an energy deposition compatible with that of an electron.
Additional selections on the event properties were applied in order to reject pileup events, as well as a selection on the longitudinal event vertex position of $|z| < 10$~cm to ensure uniform detector coverage. 
The selected data samples correspond to integrated luminosities of approximately 30 nb$^{-1}$, 1.8 pb$^{-1}$, and 8.8 pb$^{-1}$ for the MB, TRD, and HM triggers, respectively.

Several sets of Monte-Carlo (MC) simulations are used for the corrections applied in this analysis. The first simulates MB collisions and is used to correct the charged-particle multiplicity. Events
are simulated using PYTHIA 8~\cite{Bierlich:2022pfr} with Monash tune~\cite{Skands:2014pea}. Particles are propagated through the ALICE detectors using GEANT3~\cite{GEANT3}. Simulation of pileup events is also included. To account for the particle species-dependent detection efficiency, the relative abundances of pions, kaons, protons, and strange baryons in the simulation were reweighted to match those observed in the data. This reweighting follows the procedure described in  Ref.~\cite{ALICE:2022xip}.

For the simulation of the prompt and non-prompt \jpsi meson reconstruction, two MC samples are used. One has injected prompt \jpsi mesons on top of PYTHIA 6.4~\cite{Sjostrand:2006za} inelastic events. In the other, the PYTHIA 6.4 events are required to contain a $\rm b\overline{b}$ quark pair. The b quarks are hadronized to B hadrons and then forced to decay via \jpsi decay channels. The \jpsi mesons are decayed via the dielectron channel using EvtGen~\cite{Lange:2001uf}, which implements the radiative QED corrections using PHOTOS~\cite{Barberio:1993qi}. 
The events are then propagated through the ALICE detectors using GEANT3~\cite{GEANT3}.
In addition, the track parameters and their covariance matrix are smeared in order to reproduce the secondary vertexing resolution observed in the data.

%% file: C_Analysis.tex

The per-event yield of \jpsi mesons is measured as a function of the charged-particle multiplicity. Both quantities are normalized by their average values in inelastic events containing at least one charged particle within $|\eta|<1$ (an event class called INEL$>$0). The inclusive \jpsi yield is separated into prompt and non-prompt components. The separation relies on the reconstruction of the secondary decay vertex. In the case of non-prompt \jpsi mesons, it is displaced from the primary vertex by a distance on the order of hundreds of micrometers, due to the weak decay of the beauty hadron mother particles. The multiplicity is estimated in three regions with respect to the emission direction of the \jpsi meson. The analysis is performed differentially in the \jpsi transverse momentum (\pt).

The multiplicity-dependent self-normalized yields are measured separately for two event classes, namely INEL$>$0 and V0M~0--0.1\%. For the INEL$>$0 analysis, data from the MB-triggered and the TRD-triggered samples are used, and corrected for trigger efficiencies. The self-normalized yields obtained independently for both samples are combined together using, as a weight, the inverse of the quadratic sum of statistical and uncorrelated systematic uncertainties.

\subsection{Multiplicity estimator}
\label{sec:multEstimator}

The charged-particle multiplicity $N_{\rm ch}$ refers to the number of primary (i.e.\ with a proper lifetime larger than 1 cm/$c$~\cite{ALICE-PUBLIC-2017-005}) charged particles produced within $|\eta|<0.9$.
It is estimated based on the event multiplicity characterized by the number of global tracks, $N_{\rm trks}$, reconstructed both with the ITS and the TPC and with $\pt >0.15$ GeV/$c$. The track selection criteria are similar to those used in Ref.~\cite{ALICE:2018vuu}.

\begin{figure}[htb]
\centering
\includegraphics[width=0.4\linewidth]{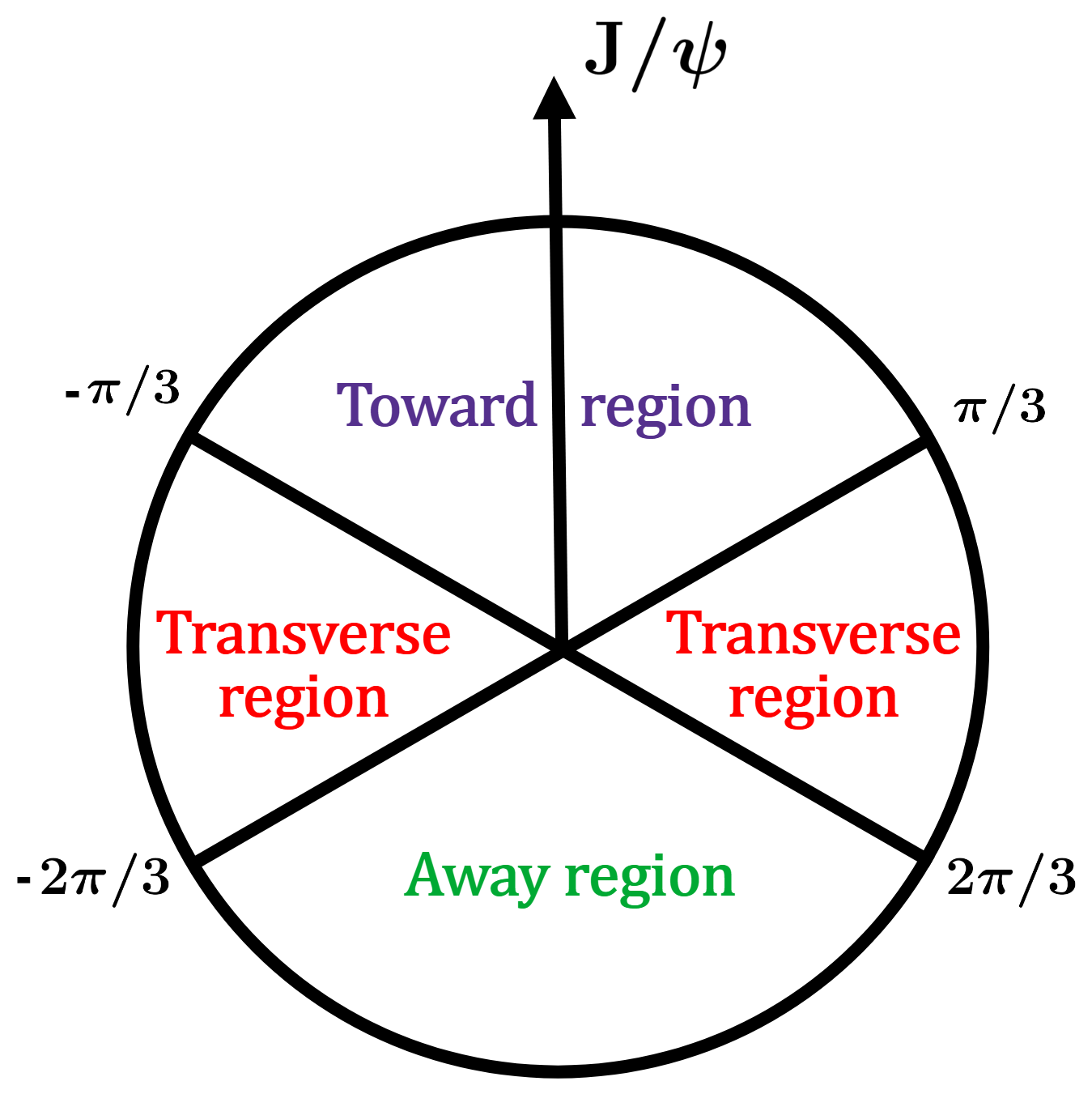}
  \caption{Regions in azimuthal angle defined with respect to a \jpsi candidate.}
    \label{fig:sketch_regions}
\end{figure}

In events containing a \jpsi candidate, the multiplicity can be separated into several azimuthal regions relative to the \jpsi emission direction. These azimuthal regions are as labelled in Fig.~\ref{fig:sketch_regions}.
The toward region contains all tracks with $|\varphi_{\rm track}-\varphi_{\Jpsi}| < \pi/3$. For the transverse region, the absolute azimuthal angle difference must be between $\pi/3$ and $2\pi/3$, while it must be larger than $2\pi/3$ for a track to be located in the away region.

In order to evaluate the multiplicity distribution of INEL$>$0 events in azimuthal regions, $N_{\rm trks,region}$, the tracks are counted in regions relative to an azimuthal angle selected randomly from a uniform distribution. Because the three regions span equal intervals in azimuth, their INEL$>$0 multiplicity distributions are identical.

\subsection{Corrections to the multiplicity}
\label{sec:multCorr}

Several corrections are applied to obtain the charged-particle multiplicity distribution from the event multiplicity. 
First, the event multiplicity distribution is corrected for the efficiency of the event selection criteria, i.e.\ the requirement of having a reconstructed vertex and the MB trigger selection. Both corrections affect mostly low-multiplicity events. The vertex reconstruction efficiency is estimated from data as a function of the event multiplicity. The MB trigger efficiency is estimated from MC simulations as a function of the V0 signal, and is further corrected to obtain an $N_{\rm trks}$-dependent efficiency using correlations between multiplicity estimators extracted from data.

An additional correction is applied to account for two effects: the inefficiency of detecting and reconstructing primary charged particles, and the remaining contamination from secondary or pileup particles. A detector response matrix, representing the correlations between the charged-particle multiplicity $N_{\rm ch}$ and the event multiplicity $N_{\rm trks}$, is extracted from the MB MC. An Iterative Bayesian Unfolding algorithm~\cite{D'Agostini:265717} uses this matrix and the $N_{\rm trks}$ distribution as an input in order to iteratively correct the full $N_{\rm ch}$ distribution. An unfolding matrix, which is an iterative correction of the detector response matrix, is also obtained at this stage.

In addition, the $N_{\rm trks}$ distribution is separated into several multiplicity intervals, in which the \jpsi yields are also measured. The $N_{\rm ch}$ distribution corresponding to each of these intervals is then determined by convoluting the $N_{\rm trks}$ distribution in this interval with the unfolding matrix. The self-normalized multiplicity value corresponding to each $N_{\rm trks}$ interval is obtained by dividing the average value from this distribution by the multiplicity-integrated $\langle N_{\rm ch}\rangle_{\rm INEL>0}$ obtained in the MB sample.  The same unfolding  procedure is applied to correct the event multiplicity distribution in azimuthal regions.

\subsection{\texorpdfstring{\jpsi}{Jpsi} selection}
\label{sec:jpsi_selection}

The \jpsi candidates are built using all possible pairs of opposite-sign tracks, following a strategy which is similar to the one reported in Refs.~\cite{ALICE:2020msa,ALICE:2021dtt}. The selected tracks are reconstructed in both the ITS and the TPC and are required to have $\pt>1$~\GeVc and $|\eta|<0.9$. A requirement of at least one hit in the SPD detector is imposed to ensure a good spatial resolution for secondary vertexing and to reject secondary electrons from photon conversions in the material outside of the SPD. In the TPC, the track candidates are required to have at least 70 clusters out of a maximum of 159 possible, which ensures good momentum and electron identification resolution. For electron identification, the specific energy loss (${\rm d}E/{\rm d}x$) in the TPC is used. The signal is required to be within 3 standard deviations (3$\sigma$) with respect to the mean expected ${\rm d}E/{\rm d}x$ for electrons, as well as more than 3$\sigma$ from the mean expected ${\rm d}E/{\rm d}x$ for both pions and protons. In the \pt-differential analysis, the latter requirement on pion and proton rejection is loosened to 2.5$\sigma$ for both tracks when the transverse momentum of the \diele pair is larger than 8 GeV/$c$.

The invariant mass ($m_{ee}$) distribution of all pair candidates between 2 and 4 GeV/$c^2$ is extracted in each multiplicity interval. A binned likelihood fit of this distribution uses three templates, for combinatorial background, correlated background and signal. The combinatorial background is modelled using an event mixing technique, where events are grouped by longitudinal position of the vertex and event multiplicity.
The normalization of the template is fixed such that the distribution of same-sign candidates from the mixed-event procedure matches the one from the same-event candidates. The correlated background is modelled by a second-order polynomial function. The shape of the signal is taken from the MC simulation with injected \jpsi signals.
The number of \jpsi counts is obtained by subtracting the two background components and counting the remaining number of candidates in the mass range from 2.92 to 3.16~GeV/$c^2$.

\subsection{Prompt and non-prompt \texorpdfstring{\jpsi}{Jpsi} separation}

In order to separate prompt and non-prompt \jpsi mesons, a Boosted Decision Tree (BDT) algorithm is used. The BDT is also used to reduce the background from uncorrelated \diele pairs, which causes the signal-to-background ratio to decrease with $N_{\rm trks}$. The BDT is trained using the ROOT TMVA package~\cite{TMVA:2007ngy}. Three classes are defined: background, prompt signal, and non-prompt signal. The sample of each class is further divided into independent training and testing samples. The background sample is taken from data, using \diele pairs from the side-bands of the invariant mass distribution ($2<\mee<2.6$ GeV/$c^2$ and $3.2<\mee<4$ GeV/$c^2$). The prompt and non-prompt samples are taken from the dedicated MC simulation with injected \jpsi mesons. The training is done with seven variables. For each daughter, the deviation of ${\rm d}E/{\rm d}x$ from the mean electron hypothesis (in number of standard deviations), the distance of closest approach (DCA) in the transverse plane, and the hit map in the SPD are used. The last used variable is  the pseudo-proper decay length of the pair, defined as

\begin{equation}
    x= \vec{L} \cdot \frac{\vec{\pt}}{\pt} \frac{c\ m_{\Jpsi}}{\pt}\, ,
\end{equation}

where $\vec{L}$ is the vector between the primary and secondary vertex, \pt is the transverse momentum of the \jpsi candidate, and $m_{\Jpsi}$ is the \jpsi mass~\cite{ParticleDataGroup:2024cfk}.
The same BDT model is used for all multiplicity intervals.

\begin{figure}[htb]
\centering
\includegraphics[width=0.6\linewidth]{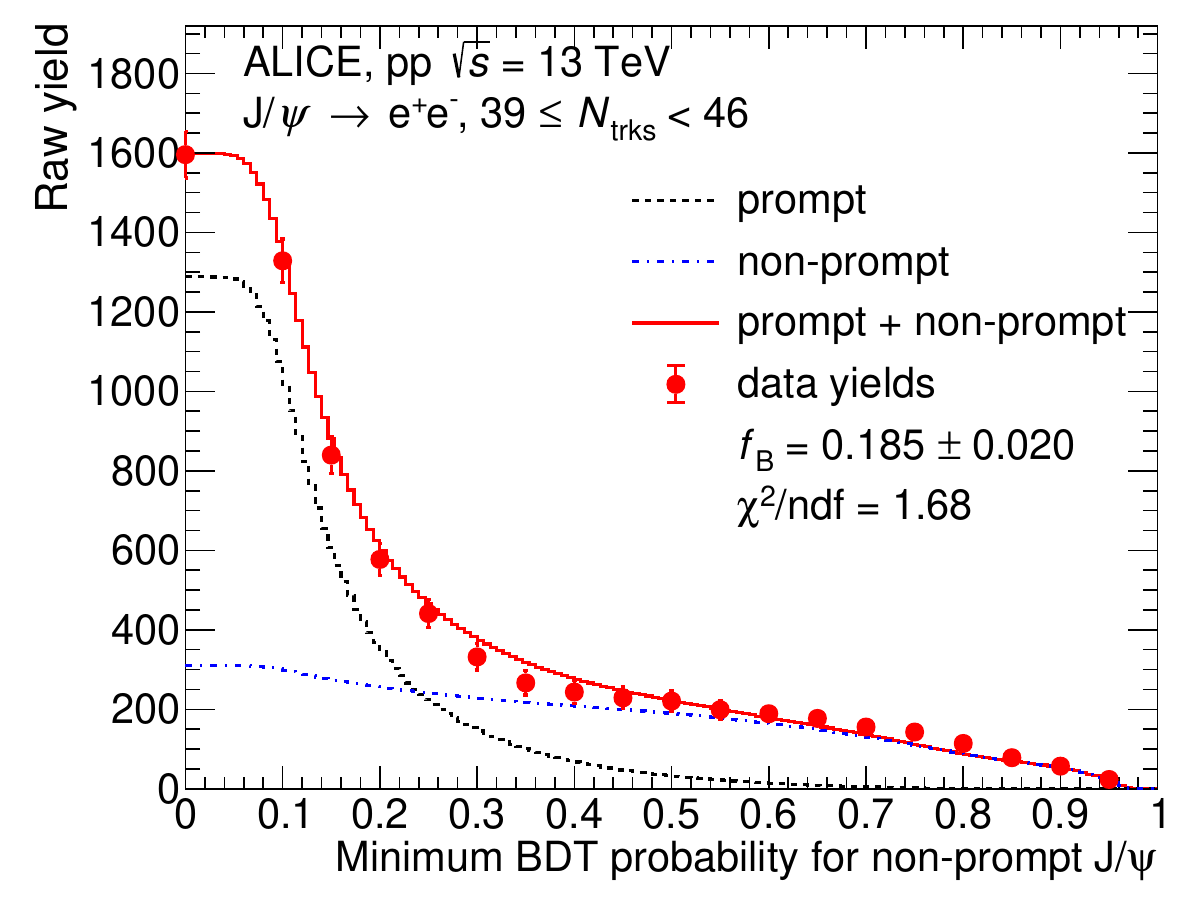}
\caption{Raw inclusive \jpsi yield as a function of the selection on the BDT non-prompt output probability for $39\leq N_{\rm trks}<46$, and $\pt >1$ GeV/$c$. The error bars indicate statistical uncertainties. The lines correspond to the prompt (dashed), non-prompt (dot-dashed) and total (continuous) \jpsi scaled efficiencies. See text for more details.}
    \label{fig_NPfractionExtraction}
\end{figure}

The output of the BDT are probabilities for a candidate to belong to each of the three classes. In order to reduce the background, a selection is done by removing candidates with high BDT output probability to be background.
The separation between prompt and non-prompt \jpsi mesons is done on a statistical basis, following a method described in detail in Ref.~\cite{ALICE:2021mgk}. The method is illustrated in Fig.~\ref{fig_NPfractionExtraction}. Using invariant mass fits, the number of raw inclusive \jpsi counts is extracted with different selections on the non-prompt BDT output probability. Each selection rejects candidates for which this output is lower than a chosen value, modifying the fraction of prompt and non-prompt components. The number of prompt and non-prompt \jpsi counts in the original sample is then extracted from a two-components fit. This fit uses as templates the BDT selection-dependent efficiencies estimated from the BDT testing sample. It consists in minimizing a $\chi^2$ which also takes into account the correlations between the number of raw counts with different selections,

This method has been compared to another method using a two-dimensional likelihood fit on the dielectron invariant mass and pseudo-proper decay length, described in detail in Ref.~\cite{ALICE:jpsi:nonpromptcspp7tev,ALICE:jpsi:nonpromptcspp13tev}. Both results agree within uncertainties.

\subsection{Corrections to the \texorpdfstring{\jpsi}{Jpsi} yields and non-prompt fraction}

\paragraph{Self-normalized \texorpdfstring{\jpsi}{Jpsi} yields:}

The self-normalized yields in the $N_{\rm trks}$ interval $i$ in each of the three triggered samples are defined as

\begin{equation}
    (N_{\Jpsi}/\langle N_{\Jpsi}\rangle_{\rm{INEL}>0})_i = \frac{N_{\Jpsi\text{, raw}}^i/ (A^i \times \varepsilon^i)}{N_{\text{evt, corr}}^i} \frac{N_{\text{evt, corr}}^{\rm{INEL}>0}}{N_{\Jpsi\text{, raw}}^{\rm{INEL}>0}/ (A^{\rm{INEL}>0} \times \varepsilon^{\rm{INEL}>0})} \, .
    \label{eq:norm_jpsi}
\end{equation}

Here, $A$ and $\varepsilon$ are the acceptance and reconstruction efficiency (including track and pair mass selections) estimated from MC, either in the specific $N_{\rm trks}$ interval (for $A^i$ and $\varepsilon^i$) or using the entire INEL$>$0 event sample (for $A^{\rm{INEL}>0}$ and $\varepsilon^{\rm{INEL}>0}$). 
$N_{\Jpsi\text{, raw}}^i$ and  $N_{\Jpsi\text{, raw}}^{\rm{INEL}>0}$ are the number of raw prompt or non-prompt \jpsi counts in the $i$th $N_{\rm trks}$  interval and the INEL$>$0 sample, respectively. The \jpsi counts are estimated from the BDT cut variation method. In addition, for TRD-triggered data, a correction is done for the \jpsi trigger efficiency, described later in this section. 

$N_{\text{evt, corr}}^i$ is the number of events in the specific $N_{\rm trks}$ interval, while $N_{\text{evt, corr}}^{\rm{INEL}>0}$ is the total number of events in the MB-triggered sample. Both quantities are corrected for vertex reconstruction efficiency, and for MB trigger efficiency in the INEL$>$0 event class.

In Eq.~\ref{eq:norm_jpsi}, $N_{\Jpsi\text{, raw}}^i$ and $N_{\text{evt, corr}}^i$ are estimated in the respective samples (MB, HM, TRD), while the INEL$>$0 quantities are estimated from the MB sample. The exception is in the TRD-triggered case, for which $N_{\Jpsi\text{, raw}}^{\rm{INEL}>0}$ is also estimated from the TRD-triggered sample and the number of events, both in $N_{\rm trks}$ intervals and INEL$>$0, is estimated from the MB-triggered sample.

\paragraph{Non-prompt fraction:}

The non-prompt fraction $f_{\rm B}$ in a given $N_{\rm trks}$ interval is estimated as the ratio between the non-prompt \jpsi yield and the sum of prompt and non-prompt yields in the same interval. All yields have been corrected for acceptance and efficiency following Eq.~\ref{eq:norm_jpsi}.

\paragraph{Corrections to the MC:}

A number of corrections are applied to this MC sample in order to improve the description of data and obtain a correct efficiency in the different multiplicity intervals.
The worse vertex resolution at lower multiplicity impacts the efficiency of the BDT. The multiplicity distributions of MC signals are therefore corrected with multiplicity-dependent weights. These weights are taken from a parametrization of the multiplicity-dependent \jpsi yield and of the non-prompt fraction obtained after the first iteration of the algorithm.
The \pt distributions also vary with multiplicity. Following a method similar to the one described in Ref.~\cite{Adamova:2017uhu}, the $N_{\rm trks}$ dependence of $\langle\pt \rangle$ of inclusive \jpsi mesons is determined. The $\langle\pt \rangle$ increases with multiplicity. A \pt-dependent weight is applied in every multiplicity interval in order to reproduce the hardening of the spectrum.
Finally, since the \pt dependence of the ratios between different beauty hadrons is not reproduced in PYTHIA 8 by the Monash tune, for non-prompt \jpsi mesons, the beauty hadron fractions in the PYTHIA 8 sample are reweighted to reproduce the Color Reconnection (CR) model Beyond Leading Color (CR-BLC) mode 2 ~\cite{Christiansen:2015yqa} process.

\paragraph{Additional corrections to the \texorpdfstring{\jpsi}{Jpsi} yields:}

Due to the selections, the decay daughters of a reconstructed \jpsi candidate are more likely to be included in the multiplicity estimator than typical primary particles. This effect artificially enhances the measured multiplicity in events containing a \jpsi candidate.
In order to remove this effect, when computing the event multiplicity for a given \jpsi candidate, each \jpsi decay daughter is counted with a probability exactly equal to the average track-level efficiency. This is done regardless of the track selection criteria the daughter may pass.

In addition, a bin-migration effect appears because a selected interval (bin) in $N_{\rm trks}$ contains a large variation in  $N_{\rm ch}$ values, with different \jpsi production yields. A correction of the measured \jpsi yields is applied so that they correspond to the yields which would be obtained with only events having the exact value of self-normalized multiplicity considered. This reduces the dependence on the choice of the $N_{\rm trks}$ intervals. This correction, typically of a few percent, depends on the variance of the $N_{\rm ch}$ distribution within an $N_{\rm trks}$ interval and on the second-order derivative of the self-normalized \jpsi-$N_{\rm ch}$ correlation. The latter is estimated from a power-law fit.
A toy model was used to validate these corrections.

For TRD-triggered data, the trigger efficiency is assumed to be the combination of two components: the self-trigger efficiency and the Underlying Event (UE) trigger efficiency. The self-trigger efficiency $\varepsilon(p_{\text{T}_{\Jpsi}})$ represents the efficiency for the event to be triggered by one of the \jpsi decay daughters. It is estimated from a data-driven method similar to the one described in Ref.~\cite{ALICE:2026htw}, where the single electron efficiency, obtained by analyzing events in the MB sample which also pass the TRD trigger requirements, is combined to a pair efficiency using \jpsi decay kinematics. The underlying event trigger efficiency $\varepsilon_{\rm UE}(N_{\rm trks})$ represents the probability for the event in which the \jpsi meson is produced to be triggered by another particle than the \jpsi meson. It is determined from data as a function of event multiplicity. It also contains the small probability to be triggered by a correlated electron from the decay of the other B hadron in the non-prompt case, estimated from PYTHIA 8 simulations. The TRD trigger efficiency $\varepsilon_{\rm TRD}$ can therefore be written as

\begin{equation}
    \varepsilon_{\rm TRD} = \varepsilon(p_{\text{T}_{\Jpsi}}) + \varepsilon_{\rm UE}(N_{\rm trks}) - \varepsilon(p_{\text{T}_{\Jpsi}})\times \varepsilon_{\rm UE}(N_{\rm trks})\, ,
\end{equation}

where the product corrects for double-counting.
The measurement is corrected for this efficiency during signal extraction by applying a weight to each \jpsi candidate equal to 1/$\varepsilon_{\rm TRD}$.

\subsection{Systematic uncertainties}

\paragraph{Self-normalized multiplicity:}
For the self-normalized multiplicity, the systematic uncertainties which are considered are: the trigger efficiency, the unfolding procedure, the MC generator, and the data-MC discrepancy in tracking. All the uncertainties are added in quadrature. 

The trigger efficiency uncertainty was reported to be 1.3\% for INEL$>$0 events~\cite{ALICE:2020msa}, and affects all intervals through the self-normalization. This value was estimated by comparing the trigger efficiency obtained from MC with an estimation which is done using data triggered by a logical OR between the V0A and the V0C. 

To estimate the systematic uncertainty of the unfolding, the number of iterations is varied, and the results are compared to the ones obtained with a different unfolding algorithm (Iterative Dynamically Stabilized~\cite{Malaescu:2009dm}). No significant differences are observed. A MC closure test further confirms the robustness of the unfolding procedure. In addition, the $N_{\rm ch}$ distribution after unfolding is folded again using the detector response matrix. The self-normalized $N_{\rm trks}$ values obtained are similar to the ones before unfolding at the per mille level. The impact of the MC statistics is checked by extrapolating the detector response matrix at high multiplicity. A difference of only 0.2\% is observed for the highest multiplicity, negligible otherwise. 

For the uncertainty due to the event generators, the results with detector response matrices obtained using PYTHIA 8 and EPOS LHC simulations are compared. This probes differences in the particle composition and \pt distribution of charged particles, as well as uncertainties due to the correction procedure of the particle composition in MC. The uncertainty reaches a maximum of 2\% (4\% for azimuthal regions) in the lowest multiplicity bin. 

To evaluate the systematic uncertainty related to tracking, the
analysis of the multiplicity distribution is repeated using track quality criterion variation. The difference in $\langle N_{\rm ch}\rangle$ with different selections, as well as ITS-TPC matching uncertainties, is used in order to estimate a tracking uncertainty for global tracks. This tracking uncertainty is applied by modifying the track reconstruction probability in MC. The analysis is repeated with this modified detector response matrix. The difference from standard results is found to be at most 0.3\%.

A summary of the systematic uncertainties is shown in Table~\ref{tab:systMult}.

\begin{table}[h]
\caption{Summary of the systematic uncertainty contributions for the estimation of the self-normalized multiplicity azimuthally-integrated and in azimuthal regions.}
\begin{center}
  \makebox[\textwidth][c]{
  \begin{tabular}{|c|c|c|}
    \hline
    Source & $N_{\rm ch}/\langle N_{\rm ch}\rangle_{\rm{INEL}>0}$ & $N_{\rm ch,region}/\langle N_{\rm ch,region}\rangle_{\rm{INEL}>0}$\\
    \hline
    MB trigger efficiency & \makebox[1.8cm][r]{1.3\%}
    & \makebox[1.8cm][r]{1.3\%}\\
    Unfolding & \makebox[1.8cm][r]{0.0 -- 0.2\%} & \makebox[1.8cm][r]{0.0 -- 0.3\%}\\
    MC generator & \makebox[1.8cm][r]{0.0 -- 2.0\%} & \makebox[1.8cm][r]{0.4 -- 4.0\%}\\
    Tracking efficiency & \makebox[1.8cm][r]{0.1 -- 0.3\%}  & \makebox[1.8cm][r]{0.1 -- 0.2\%} \\
    \hline
    Total & \makebox[1.8cm][r]{1.3 -- 2.4\%} & \makebox[1.8cm][r]{1.4 -- 4.2\%}\\
    \hline
  \end{tabular}
  }
  
  \label{tab:systMult}
\end{center}
\end{table}

\paragraph{Non-prompt fraction and self-normalized \texorpdfstring{\jpsi}{Jpsi} yields:}
For the non-prompt fraction and self-normalized yields, the systematic uncertainties which are considered are: the trigger efficiency, the unfolding procedure, the signal extraction, the BDT training, the BDT selection cuts, the simulated \pt spectrum of \jpsi production, the beauty hadron compositions, the primary vertex calculation, as well as the possibly modified tracking efficiency in events with a \jpsi meson compared to unbiased events. Unless stated otherwise, all uncertainties are calculated by evaluating the non-prompt fraction as well as the self-normalized prompt and non-prompt \jpsi yields for each of the variations, and taking the RMS.

The efficiency of the MB trigger can affect the \jpsi yields through both the number of inelastic events and the number of \jpsi counts. For the former case, an uncertainty due to this effect was already assigned to the self-normalized multiplicity. Because the trend followed by the data is little affected by variations of this efficiency, no additional uncertainty is assigned for the yields. For the latter case, some \jpsi mesons might escape the trigger selection. Their number was estimated through the MB MC simulations. It was found to be below 2\% in the lowest multiplicity interval, and below 0.2\% otherwise.
For the TRD trigger efficiency, the uncertainty of the self-trigger efficiency mostly cancels in the self-normalization. The UE efficiency has also been varied by $\pm 15\%$, which is the discrepancy between estimated and true UE efficiency observed in an independent PYTHIA 8 standalone  simulation.

The uncertainty related to the correction for bin migration is evaluated by comparing the correction obtained when using a power-law fit and the one using a second-order polynomial fit, both with a null value at the origin and with a small offset. 

The uncertainty due to signal extraction is estimated by changing the fit range, the signal mass window, and the background shape. For the latter, the combinatorial background is changed from mixed-event to like-sign estimation, and the correlated background shape is varied from a second-order polynomial to an exponential.

For the BDT training uncertainty, the analysis is repeated with different training parameters. The separation between the training and testing samples is also randomly modified twenty times. 
Modification of the BDT selections provides a check on the reproduction in the MC of the variables used for the BDT, as well as on the instabilities in the $\chi^2$ minimization procedure. The rejection value on the BDT background output probability is modified, as well as the number and values of selections in the BDT non-prompt output probability.

A systematic uncertainty might arise from possible differences between the real and simulated \jpsi \pt spectra. To estimate the impact of these differences, different \pt distributions are used for calculating the acceptance, reconstruction efficiency, and BDT selection efficiency.  The default prompt \pt shape in MB events is based on a fit to prompt cross section data at $\sqrt{s} = 13$ TeV~\cite{ALICE:jpsi:nonpromptcspp13tev}, while its variation is taken from PYTHIA 8 with the Monash tune~\cite{Skands:2014pea}. The default non-prompt \pt shape in MB events is taken from PYTHIA 8 with Monash tune, while its variation is obtained from a Fixed Order plus Next-to-Leading Logarithm (FONLL)~\cite{Cacciari:1998it, Cacciari:2012ny} calculation. The evolution of $\langle\pt\rangle$ with multiplicity is, as a variation, taken from PYTHIA 8 (Monash tune with oniaShower settings~\cite{Cooke:2023ldt}) for prompt and non-prompt \jpsi production independently. The beauty hadron fractions, assumed by default to be independent on the multiplicity, are also varied using the multiplicity dependence observed in PYTHIA 8 with CR-BLC mode 2.

An uncertainty for the description of the primary vertex is evaluated by removing the \jpsi candidate decay daughters from the primary vertex calculation, and computing the difference to the default variation, affecting mainly the lowest multiplicities.

Finally, the reconstruction efficiency for tracks in the multiplicity estimator could be different in events containing a \jpsi meson compared to MB events. For the non-prompt \jpsi yields, the effect of a few additional non-prompt tracks with lower selection efficiency was studied using a toy model. This toy model was also used to estimate the error coming from the method for counting the \jpsi decay daughters, when assuming that there is a mismatch of 3\% between the true tracking efficiency and the one used in the reconstruction. For the measurement in azimuthal regions, the track reconstruction, \jpsi selection, and TRD trigger efficiency depend slightly on the azimuthal angle. The variation of the efficiency in each of the three regions of $|\varphi_{\rm track} - \varphi_{\jpsi}|$  compared to MB events does not exceed 0.3\%. This was also converted to an uncertainty on the multiplicity-dependent \jpsi yields using a toy model.

The systematic uncertainties from all the sources described here are assumed to be uncorrelated and added in quadrature. A typical example of the systematic contributions for prompt and non-prompt self-normalized yields, as well as for $f_{\rm{B}}$, for two intervals of $N_{\rm trks}$, is shown in Table~\ref{tab:syst}.

\begin{table}[h]
\caption{Example of systematic uncertainty contributions for the prompt and non-prompt self-normalized \jpsi yields, and for the non-prompt fraction $\rm f_{B}$. The uncertainties are shown for $\pt>1$~\GeVc, in the interval $15 \leq N_{\rm trks} \leq 19$ for INEL$>$0 sample (noted as (1)), and in the interval $39 \leq N_{\rm trks} \leq 45$ for HM sample (noted as (2)).}
\begin{center}
  \makebox[\textwidth][c]{
  \begin{tabular}{|c|c|c|c|c|c|c|}
    \hline
     Source & \multicolumn{2}{c|}{Prompt \jpsi} & \multicolumn{2}{c|}{Non-prompt \jpsi} & \multicolumn{2}{c|}{$\rm f_{B}$} \\ 
     & (1) & (2) & (1) & (2) & (1) & (2) \\ \hline
    MB trigger & 0.2\% & 0.2\% & 0.1\% & 0.1\%  & 0.0\% & 0.0\%\\
    Bin migration & 1.1\% & 0.7\% & 0.9\% & 0.6\% & 0.3\% & 0.1\% \\
    Signal extraction & 0.7\% & 2.9\% & 2.2\% & 2.1\% & 2.8\% & 4.5\% \\
    BDT training & 0.3\% & 1.6\% & 1.7\% & 5.1\% & 1.4\% & 2.6\%\\
    BDT selections & 0.5\% & 1.4\%  & 1.6\% & 5.5\% & 2.6\% & 3.0\% \\
    \pt distribution & 1.6\% & 0.3\% & 0.5\% & 0.5\% & 2.3\% & 1.9\% \\
    B hadrons composition & 0.0\% & 0.1\% & 0.2\% & 0.7\% & 0.2\% & 0.7\% \\
    Vertex reconstruction & 0.2\% & 0.3\% & 0.8\% & 0.0\% & 0.6\% & 0.0\% \\
    TRD trigger efficiency & 0.1\% & -- & 0.0\% & -- & 0.3\% & -- \\
    Tracking eff. in \jpsi events & 0.3\% & 0.0\% & 1.3\% & 1.0\% & 1.6\%& 1.3\% \\
    \hline
    Total & 2.2\% & 3.8\% & 4.5\% & 8.0\% & 5.1\% & 5.9\%\\
    \hline
  \end{tabular} 
  \label{tab:syst}
}
\end{center}

\end{table}

%% file: D_Results.tex
\subsection{Multiplicity-dependent prompt and non-prompt \texorpdfstring{\jpsi}{Jpsi} yields}
\label{sec:resultsYields}

\begin{figure}[!htb]
  \centering
    \includegraphics[width=0.8\linewidth]{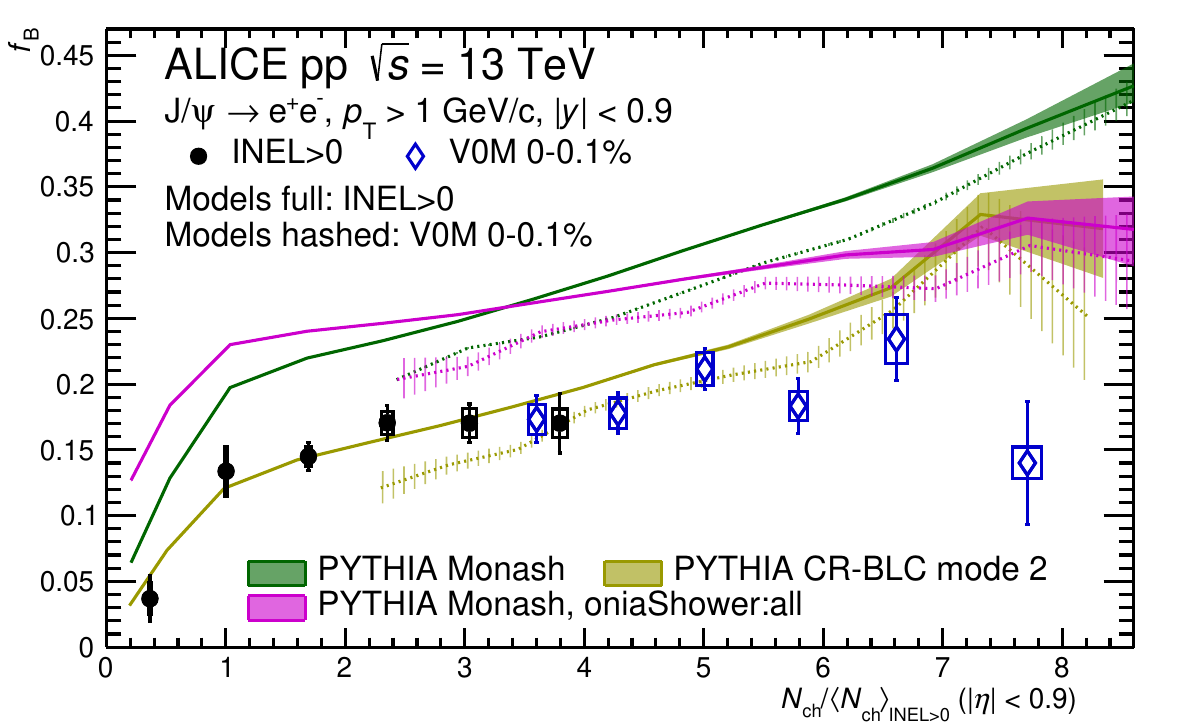}
  \caption{Fraction of non-prompt \jpsi mesons, $f_{\rm B}$, for $\pt>1$~\GeVc as a function of the self-normalized charged-particle multiplicity measured within $|\eta|<0.9$ for the INEL$>$0 (full markers) and V0M~0--0.1\% (open markers) event classes. The vertical bars and boxes indicate the statistical and systematic uncertainties, respectively. 
  The results are compared to PYTHIA 8 simulations using different tunes, as indicated by the legend~\cite{Bierlich:2022pfr, Christiansen:2015yqa, Cooke:2023ldt}. The PYTHIA calculations emulating the V0M~0--0.1\% class (0.1\% highest multiplicity events in the V0 acceptance) are shown as hashed lines.}
    \label{fig_results_fB}
\end{figure}

The fraction of \jpsi mesons with $\pt > 1$~\GeVc coming from the decay of beauty hadrons, $f_{\rm B}$, is shown in Fig.~\ref{fig_results_fB} as a function of self-normalized multiplicity, for both the INEL$>$0 (full markers) and V0M~0--0.1\% (open symbols) data samples.  For all the measurements presented in this section, the decay products of the \jpsi meson are kept in the multiplicity count. A more thorough discussion on this choice can be found in App.~\ref{sec:daughters}. This fraction provides insight into the difference between charm and beauty production mechanisms.
In the lowest multiplicity interval, $f_{\rm B}$ has a relatively small value compared to higher event multiplicities. This is expected due to the larger associated multiplicity expected in beauty-hadron decays compared to prompt \jpsi decays. For higher multiplicities, a linear fit, excluding the first multiplicity interval, deviates from a flat trend by 2.9$\sigma$, suggesting a hint of an increase with multiplicity.

The data are compared to PYTHIA 8.311 simulations using several different settings: the Monash tune~\cite{Skands:2014pea}, the CR-BLC mode 2 process~\cite{Christiansen:2015yqa}, and the oniaShower process~\cite{Cooke:2023ldt} with the Monash tune. PYTHIA is a MC generator which simulates sequentially multiple partonic interactions. These can further be evolved through partonic showers. Hadronization occurs via the breaking of strings between quarks and gluons. The CR mechanism allows the strings to be reorganized in order to minimize the string length.
CR is improved with the CR-BLC mechanisms, which, by rearranging the color flow in the event, could strongly impact the multiplicity. Prompt quarkonia are produced by using the NRQCD framework~\cite{Baier:1983va, Cho:1995ce, Nason:1999ta}, typically in the initial hard scattering. When the oniaShower process is activated, quarkonia are produced within partonic showers, using splitting kernels for the emission of heavy quark-antiquark pairs.

The data in the next figures are also compared to EPOS4HQ1.0~\cite{Werner:2023zvo, Werner:2023fne, Werner:2023mod, Werner:2023jps}, an event generator implementing parallel scatterings between projectile and target nucleons. A dynamical saturation scale, dependent on the momentum fraction but also on the number of such scatterings, is introduced in the calculation in order to ensure that the factorization theorem holds and to model non-linear effects. The partonic ladders from the scatterings are further evolved in a core-corona procedure. The high energy-density core evolves hydrodynamically while the lower energy-density corona directly enters hadronization. The evolution of the core can also be deactivated. Heavy quarks are created in the partonic ladder and interact in the medium via elastic and radiative collisions, then hadronize via coalescence or fragmentation~\cite{Zhao:2023ucp, Zhao:2024ecc}.  If a charm and an anticharm quark are close in position space and momentum, they can also hadronize together by coalescence to form a quarkonium. The quarks are assumed to be initially closer in position space when they come from the same heavy-quark pair~\cite{Zhao:2025cnp}. The fill areas for the PYTHIA and EPOS4 curves represent statistical uncertainties. They are sizeable at high multiplicity especially for EPOS4 with hydrodynamic evolution.

The prompt \jpsi results are further compared to two CGC-based calculations. The first one, uses the CGC initial-state condition together with the Improved Color Evaporation Model (CGC+ICEM)~\cite{Ma:2018bax}. High-multiplicity events are modelled with a higher saturation scale, modifying the gluon distribution inside the proton. A second calculation employs the 3-Pomeron Color Glass Condensate model (3-Pomeron CGC)~\cite{Siddikov:2019xvf, Gotsman:2020ubn}. The 3-Pomeron fusion mechanism enhances the associated multiplicity compared to 2-Pomeron fusion, by increasing the possible fluctuations in the multiplicity produced by each Pomeron.

The self-normalized multiplicity distributions are well reproduced by the models, with a discrepancy of no more than 20\% until at least 5 times the average multiplicity, and a larger discrepancy for higher multiplicities. In order to compare the model calculations to the ALICE HM-triggered data, a similar selection as for the data is performed, namely for the 0.1\% events with the highest charged-particle multiplicity in the acceptance of the V0 detector. The model calculations with these selections are shown as hashed curves in the figures. An inspection on the MC revealed a small impact of the V0 detector effects on the correlation compared to the data uncertainties, due to the requirement in the models not being applied on the V0M signal but on the multiplicity in the V0 acceptance. According to the MC simulations based on these models, the HM trigger selection induces a bias on the measured $f_{\rm B}$ at a given event multiplicity compared to the same observable in INEL$>$0 events. Such a bias has also been observed when analyzing jet production~\cite{ALICE:2023plt}. In a naive scenario in which the high multiplicity requirement at forward rapidity implicitly selects events with enhanced beauty-quark pair production at forward rapidity, a reduction of $f_{\rm B}$ at midrapidity is expected.
Overall, the model calculations overestimate the non-prompt fraction, with the exception of the CR-BLC PYTHIA settings.

\begin{figure}[!htb]
  \centering
    \includegraphics[width=0.95\linewidth]{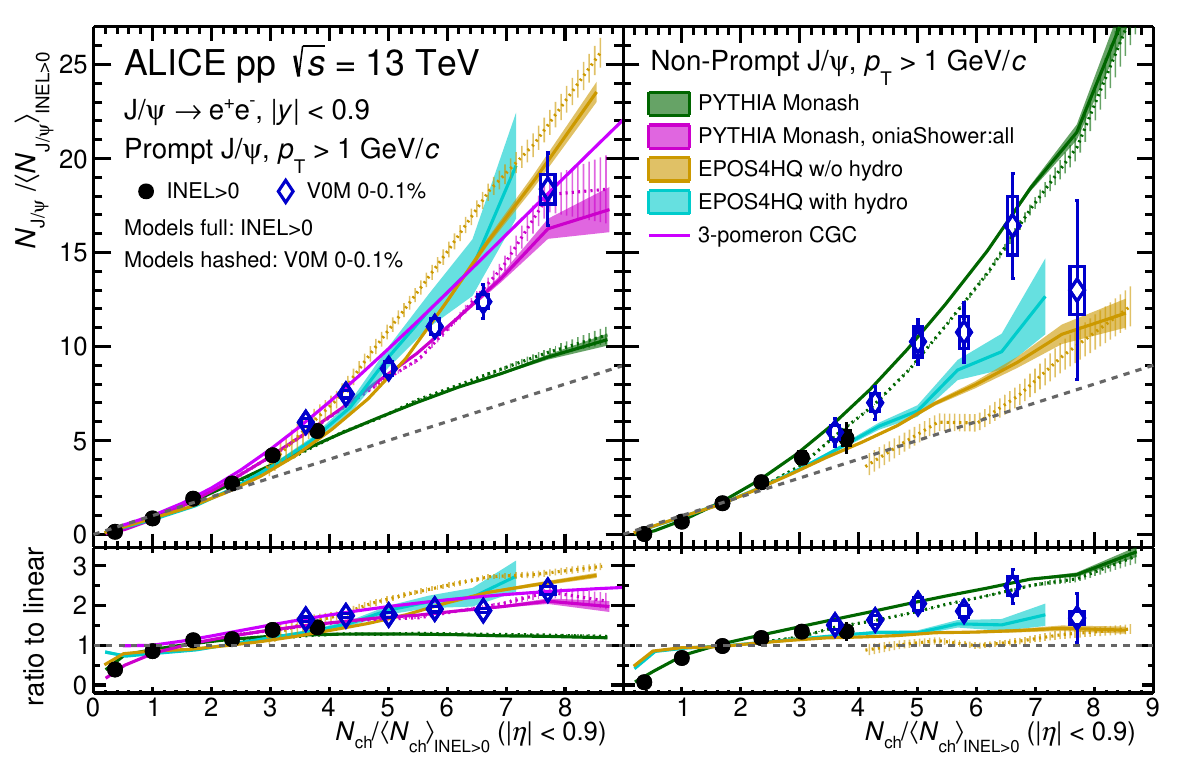}
  \caption{Self-normalized prompt (left) and non-prompt (right) \jpsi yields with $\pt>1$~\GeVc as a function of self-normalized charged-particle multiplicity at midrapidity. The data are shown separately for the INEL$>$0 and V0M~0--0.1\% event classes. The yields are compared to PYTHIA 8  using different settings~\cite{Skands:2014pea, Cooke:2023ldt}, as well as EPOS4~\cite{Werner:2023zvo, Werner:2023fne, Werner:2023mod, Werner:2023jps} and the 3-Pomeron CGC model~\cite{Siddikov:2019xvf, Gotsman:2020ubn}. Lower panels show the ratios to the expectation from a linear increase. The vertical bars and boxes indicate the statistical and systematic uncertainties, respectively.}
\label{fig_results_prompt_nonprompt}
\end{figure}

In the upper panels of Fig.~\ref{fig_results_prompt_nonprompt}, the self-normalized yields of prompt and non-prompt \jpsi mesons are shown as a function of the charged-particle multiplicity. Both the \jpsi yields and the multiplicity are normalized by their average value in INEL$>$0 events, see Eq.~\ref{eq:norm_jpsi}. 
A dashed grey line representing a linear increase with slope 1 ($y=x$) is also drawn for reference and illustration purposes. The bottom panels show the ratio between the data points and the linear reference.
Both prompt and non-prompt \jpsi yields show a stronger-than-linear increase with multiplicity, similar to the one found for inclusive \jpsi yield~\cite{ALICE:2020msa}. The increase of \jpsi yield with multiplicity is similar for the prompt and non-prompt components, as expected from the mild dependence of the $f_{\rm B}$ with multiplicity.

The stronger-than-linear increase could be interpreted as a larger saturation for soft particle production than for hard particle production in high-density environments, or as an increase in the associated particle production for hard probes.
Comparisons of the measurements to PYTHIA calculations show that the Monash tune underestimates the prompt \jpsi yield for multiplicities higher than four times the average multiplicity. The CR-BLC mode 2 settings, shown separately in App.~\ref{sec:PythiaAdditional} for visibility, present a similar increase to the Monash tune. In contrast, the oniaShower option is in very good agreement with the data. This could indicate that, due to the additional particles emitted in the parton shower, the production process of the \jpsi meson has a strong influence on the multiplicity-dependent \jpsi yield, and that higher-order corrections are important to describe quarkonium production. 
For non-prompt \jpsi production, the Monash tune reproduces the trend as a function of multiplicity well. The oniaShower process is not shown since it only affects prompt \jpsi production. As was also seen for $f_{\rm B}$, PYTHIA calculations predict that, at a given multiplicity, the self-normalized yield obtained for non-prompt \jpsi production is lower in HM-triggered events than in INEL$>$0 events. The prompt \jpsi production seems less affected by this trigger bias, as estimated from PYTHIA.
The 3-Pomeron CGC calculation shows a good agreement to the prompt \jpsi data over the entire multiplicity range. The EPOS4 calculations, with and without the hydrodynamic evolution, describe the prompt \jpsi measurements fairly well, except for a tendency to slightly overshoot the V0M~0--0.1\% results. The strong increase in this case is mainly due to the combinatorial enhancement from c and $\overline{\rm c}$ coming from different $\rm c\overline{c}$ pairs, more frequent at higher multiplicities~\cite{Zhao:2025cnp}. A slightly stronger increase is found with hydrodynamic evolution compared to without, due to a reduction in the multiplicity for which the available energy has been transferred to radial flow~\cite{Werner:2023zvo}. EPOS4 calculations underestimate the non-prompt \jpsi measurements for a self-normalized multiplicity above 3.

\begin{figure}[!htbp]
  \centering
    \includegraphics[width=1.\linewidth]{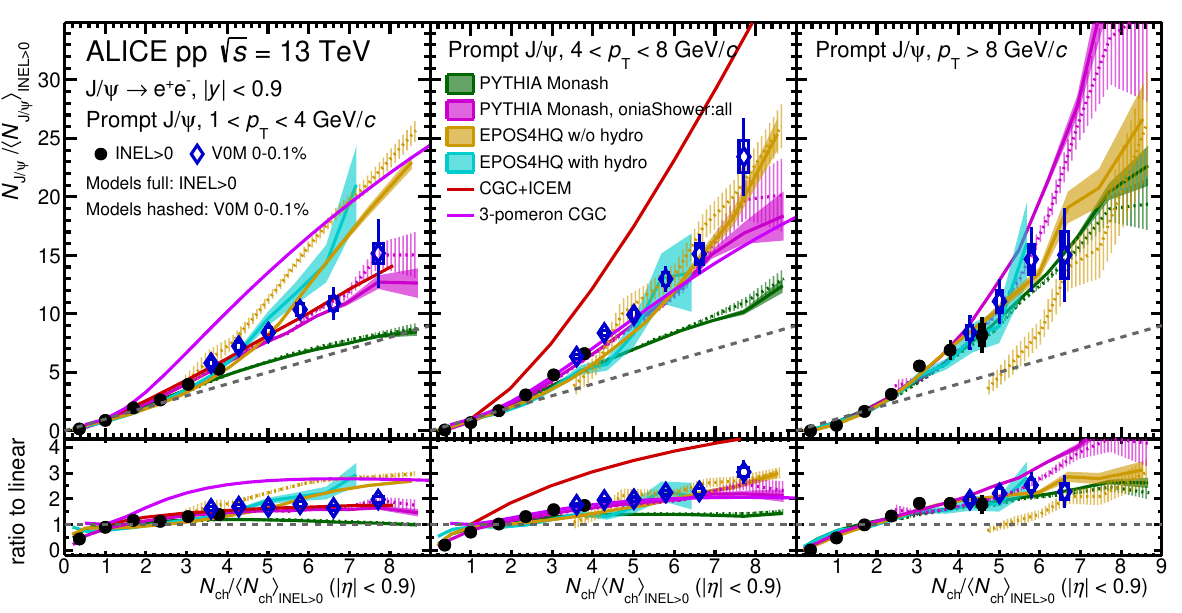}
  \caption{Prompt \jpsi yield for $1<\pt<4$~\GeVc (left), $4<\pt<8$~\GeVc (middle) and $\pt>8$~\GeVc (right) as a function of self-normalized charged-particle multiplicity within $|\eta|<0.9$ for the INEL$>$0 and V0M~0--0.1\% event classes. The data are compared to PYTHIA 8  using different settings~\cite{Skands:2014pea, Cooke:2023ldt}, as well as EPOS4HQ~\cite{Werner:2023zvo, Werner:2023fne, Werner:2023mod, Werner:2023jps}, the 3-Pomeron CGC model~\cite{Siddikov:2019xvf, Gotsman:2020ubn} and the CGC+ICEM model~\cite{Ma:2018bax}. Lower panels show the ratios to the expectation from a linear increase. The vertical bars and boxes indicate the statistical and systematic uncertainties, respectively.
  \vspace{0.1cm}
  }
    \label{fig_results_prompt_vspt}
\end{figure}

\begin{figure}[!htb]
  \centering
    \includegraphics[width=1.\linewidth]{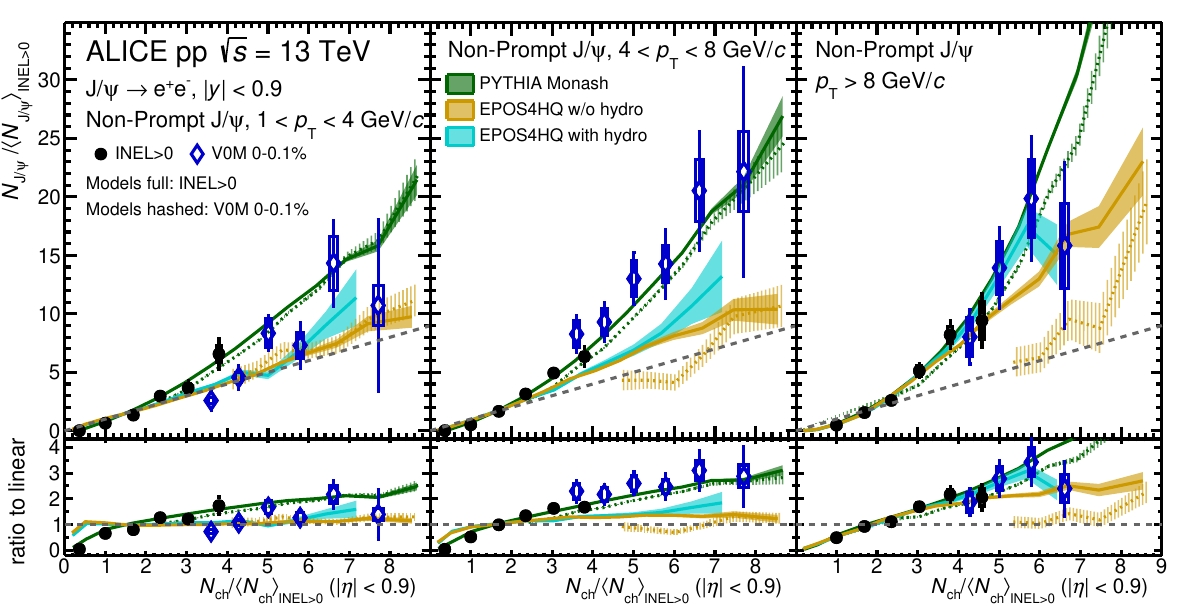}
  \caption{
  Non-prompt \jpsi yield for $1<\pt<4$~\GeVc (left), $4<\pt<8$~\GeVc (middle) and $\pt>8$~\GeVc (right) as a function of self-normalized charged-particle multiplicity within $|\eta|<0.9$ for the INEL$>$0 and V0M~0--0.1\% event classes. The data are compared to PYTHIA 8 using different settings~\cite{Skands:2014pea, Cooke:2023ldt}, as well as EPOS4HQ with and without hydrodynamics activated~\cite{Werner:2023zvo, Zhao:2023ucp}. Lower panels show the ratios to the expectation from a linear increase. The vertical bars and boxes indicate the statistical and systematic uncertainties, respectively. 
  \vspace{0.1cm}
  }
    \label{fig_results_nonprompt_vspt}
\end{figure}

Figures~\ref{fig_results_prompt_vspt} and~\ref{fig_results_nonprompt_vspt} show the multiplicity dependence of the self-normalized yields separately in three \pt intervals ($1<\pt<4$, $4<\pt<8$, and $\pt>8$~\GeVc) for prompt and non-prompt \jpsi mesons, respectively. In both cases,  the slope of the multiplicity dependence is growing with increasing \pt, as was previously seen also for inclusive \jpsi mesons~\cite{ALICE:2020msa}. For all the \pt intervals, the slope is larger than 1. The increase of the slope is more pronounced between the first and second intervals than between the second and third. The HM trigger bias also seems noticeable in the data for non-prompt \jpsi production in the lowest \pt interval. 

The comparison to PYTHIA and EPOS4 calculations is qualitatively similar to the one for \pt-integrated yields in the lowest two \pt intervals, with the exception of prompt \jpsi production in the EPOS4 calculations. In the latter, no strong modification between low- and high-\pt \jpsi yields is observed, leading to an overestimation at high multiplicity and low momentum. This could be explained by the more abundant number of charm quarks compared to higher momentum, which results in a larger combinatorial contribution. For $\pt>8$~\GeVc, all PYTHIA and EPOS4 calculations reproduce the data, although uncertainties are larger. The CGC-based calculations describe the data only partially, with a good description of the $1<\pt<4$~\GeVc results for the CGC+ICEM and of the $4<\pt<8$~\GeVc results for the 3-Pomeron implementation. For the other \pt intervals, these calculations either do not provide results or are in disagreement with the data.

\subsection{Results as a function of multiplicity in azimuthal regions}

\begin{figure}[!htb]
  \centering
    \includegraphics[width=0.9\linewidth]{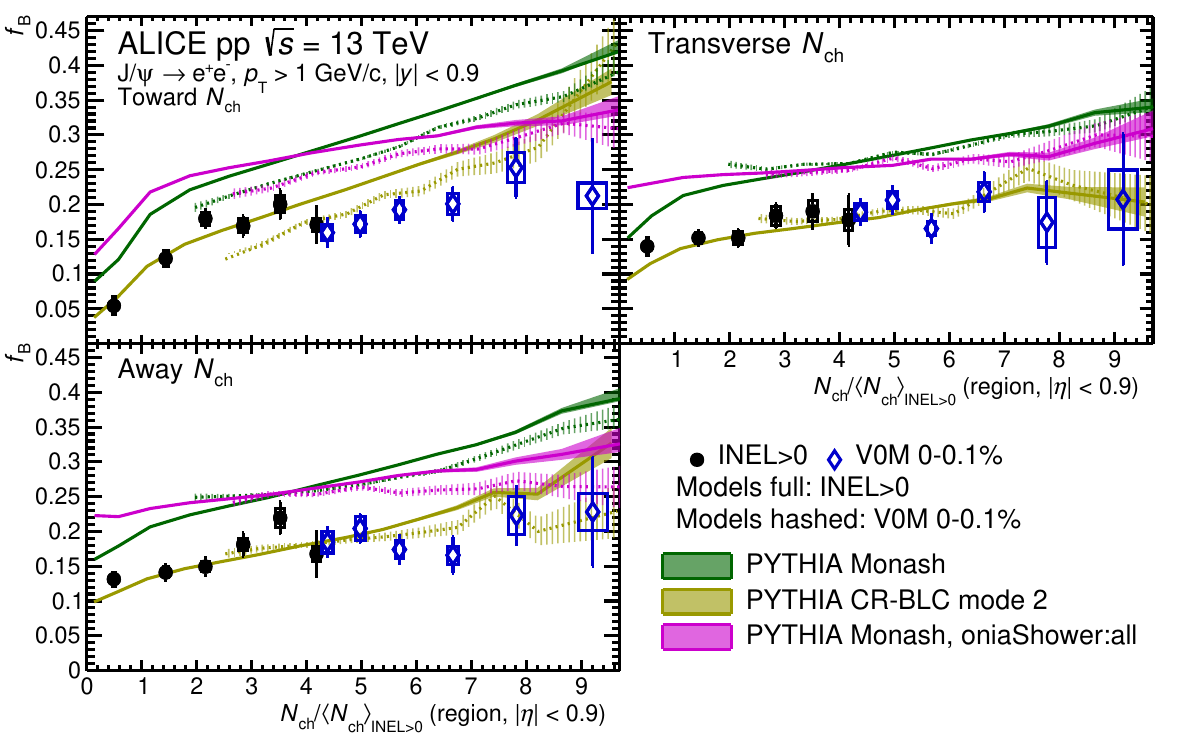}
  \caption{The fractions of \jpsi mesons from beauty feed-down with $\pt>1$~\GeVc as a function of the self-normalized charged-particle multiplicity in $|\eta|<0.9$ in the toward (top-left), transverse (top-right), and away (bottom left) azimuthal regions with respect to the \jpsi momentum direction. 
  The data are shown separately for the INEL$>$0 and V0M~0--0.1\% event classes and are compared to results from PYTHIA 8 simulations using different settings~\cite{Skands:2014pea, Christiansen:2015yqa, Cooke:2023ldt}. The vertical bars and boxes indicate the statistical and systematic uncertainties, respectively. }
    \label{fig_results_fB_regions}
\end{figure}

In Fig.~\ref{fig_results_fB_regions}, the fractions of non-prompt \jpsi mesons with $\pt>1$~\GeVc are shown as a function of the self-normalized multiplicity measured in the toward (top-left), transverse (top-right) and away (bottom-left) azimuthal angle regions as defined in Sec.~\ref{sec:multEstimator}.
All regions present a hint of an increase of the non-prompt \jpsi fraction as a function of multiplicity. A linear fit where the first point is excluded gives a significance compared to a flat trend of 2.9$\sigma$ (2.5$\sigma$ and 2.4$\sigma$)  for the toward (transverse and away) region, respectively. 

The PYTHIA calculations indicate a hierarchy of slopes for the increase of $f_{\rm B}$ with the event activity. It is the strongest in the toward region and the weakest in the transverse region. 
They also indicate an HM-trigger bias in the measurement using the toward multiplicity estimator comparable to the one for inclusive multiplicity, while this trigger bias is smaller in the transverse and away regions.

\begin{figure}[!htb]
  \centering
    \includegraphics[width=1.\linewidth]{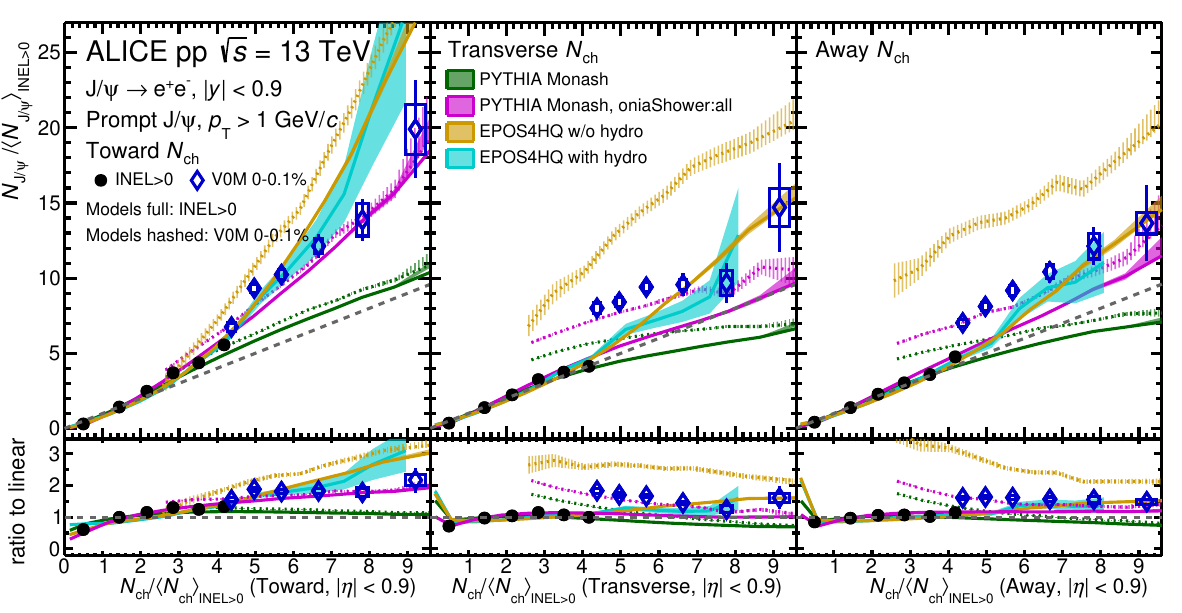}
  \caption{Self-normalized prompt \jpsi yields with $\pt>1$~\GeVc as a function of the self-normalized charged-particle multiplicity in $|\eta|<0.9$ measured in the toward (left), transverse (middle), and away (right) azimuthal regions with respect to the \jpsi momentum direction. 
  The data are shown separately for the INEL$>$0 and V0M~0--0.1\% event classes and are compared to PYTHIA 8 simulations using different settings~\cite{Skands:2014pea, Cooke:2023ldt}, as well as EPOS4HQ with and without hydrodynamics activated~\cite{Werner:2023zvo, Zhao:2023ucp}. The lower panels show the ratios to the expectation from a linear increase. The vertical bars and boxes indicate the statistical and systematic uncertainties, respectively.}
    \label{fig_results_prompt_regions}
\end{figure}

\begin{figure}[!htb]
  \centering
    \includegraphics[width=1.\linewidth]{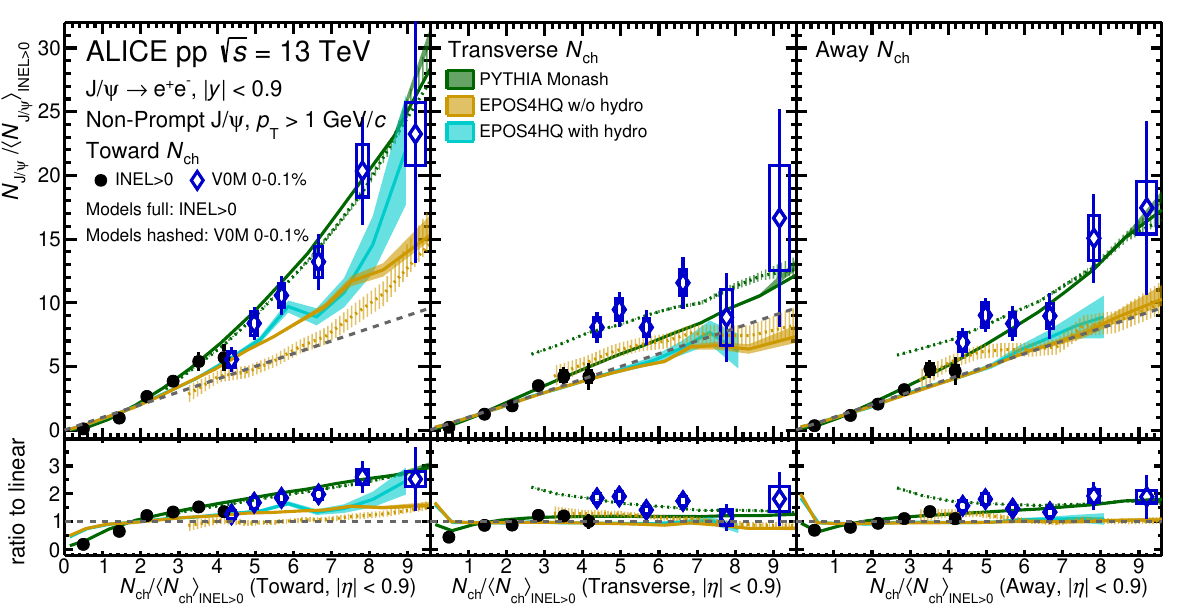}
  \caption{
  Self-normalized non-prompt \jpsi yields with $\pt>1$~\GeVc as a function of the self-normalized charged-particle multiplicity in $|\eta|<0.9$ measured in the toward (left), transverse (middle), and away (right) azimuthal regions with respect to the \jpsi momentum direction. 
  The data are shown separately for the INEL$>$0 and V0M~0--0.1\% event classes and are compared to PYTHIA 8 simulations using different settings~\cite{Skands:2014pea, Cooke:2023ldt}, as well as EPOS4HQ with and without hydrodynamics activated~\cite{Werner:2023zvo, Zhao:2023ucp}. Lower panels show the ratios to the expectation from a linear increase. The vertical bars and boxes indicate the statistical and systematic uncertainties, respectively. 
  }
    \label{fig_results_nonprompt_regions}
\end{figure}

Figures~\ref{fig_results_prompt_regions} and ~\ref{fig_results_nonprompt_regions} show the self-normalized yields as a function of the event activity in the three azimuthal angle regions for the prompt and non-prompt \jpsi yields, respectively. In both cases, the increase with multiplicity is significantly stronger for the toward region compared to the other regions. This is presumably due to an autocorrelation with the particles produced along the \jpsi direction in the same process. For the transverse and away multiplicity estimators, the \jpsi yields at high multiplicity are only slightly higher than the $y=x$ reference.

The high-multiplicity trigger introduces a strong bias in the transverse and away multiplicities.  This HM trigger bias is less significant in the toward region. It was pointed out that this trigger enhances multi-jet events with at least one jet in the V0 acceptance~\cite{ALICE:2023plt}. Therefore, at a given multiplicity in one region, the presence of a trigger could increase the multiplicity in the other regions, as was observed in the PYTHIA simulations. For the transverse and away regions, the toward-region multiplicity is increased, which could therefore increase the probability to find a \jpsi meson in this region. This trigger bias, making the interpretation of the results harder, is stronger at lower multiplicity than at higher multiplicity, as can be seen also in the model curves. Since the \jpsi meson is in the toward region, if the effect would come from an increased toward-region multiplicity, comparing INEL$>$0 and V0M~0--0.1\% event classes at a similar toward multiplicity might not allow for a large HM trigger bias to appear. 

Larger Poissonian fluctuations are expected for $N_{\rm ch,\ region}$ than for $N_{\rm ch}$, enhancing the number of events with large self-normalized $N_{\rm ch,\ region}$ compared to large self-normalized $N_{\rm ch}$. Therefore, the average $N_{\rm ch}/\langle N_{\rm ch}\rangle$ value in events where $N_{\rm ch,\ region}/ \langle N_{\rm ch,\ region} \rangle$ is chosen as a given value $n_0$ is lower than $n_0$. Thus, measurements of the multiplicity in azimuthal regions, while the \jpsi meson is measured in the full azimuth, cause in general a weaker increase compared to measurements using inclusive multiplicity. Therefore, the increase of the \jpsi yields as a function of the transverse and away multiplicity could be stronger than the baseline of soft particle production. This is also confirmed by PYTHIA simulations, which predict that the increase of pion yields with the transverse and away multiplicity is weaker than linear. These PYTHIA simulations also show that, even for pions, the increase is stronger in the toward region compared to other regions. Therefore, the difference measured between the azimuthal regions could be partly caused by the definition of these regions and not only by additional autocorrelation effects for hard particle production.

Similar to the inclusive multiplicity case, PYTHIA Monash reproduces non-prompt \jpsi yields in all regions while the oniaShower setting is necessary for reproducing the prompt yields. A stronger increase in the prompt yield as a function of toward multiplicity can be expected in oniaShower due to a modification of the production process and to the particles emitted in the parton shower. With the oniaShower process turned on, a stronger increase is also observed in the transverse region, possibly due to large angle radiations  in the partonic shower. The transverse region might not be completely free of autocorrelations, even though they are smaller than in the other regions. In addition, oniaShower also predicts a stronger increase than Monash for the away region, which could be due to back-to-back jets. Contrary to the prompt case, for non-prompt \jpsi production Monash predicts a stronger increase in the away region compared to the transverse region, possibly due to $\rm b\overline{b}$ correlations. However, here, the statistics do not allow a strong conclusion to be made on whether this is also the case in non-prompt data. At low \pt, when the \jpsi meson is not boosted, the \jpsi decay daughters can also be included in other regions than the one towards the \jpsi meson emission direction. 

For the EPOS4 calculations, the prompt yield seems to be overestimated at high multiplicity in the toward region when hydrodynamic evolution is not activated. There is also a large difference when considering the V0M~0--0.1\% event class, and the increase is slightly stronger for prompt \jpsi production without hydrodynamics. For the transverse and away regions, EPOS4 also predicts a strong increase for prompt \jpsi production as well as a strong trigger bias which largely overestimates the high-multiplicity data. In the non-prompt case, EPOS4 consistently underestimates the high-multiplicity yields for all regions. The hydrodynamic effect increases the correlation significantly only at high multiplicity in the toward region.

\begin{figure}[htb]
  \centering
    \includegraphics[width=1.\linewidth]{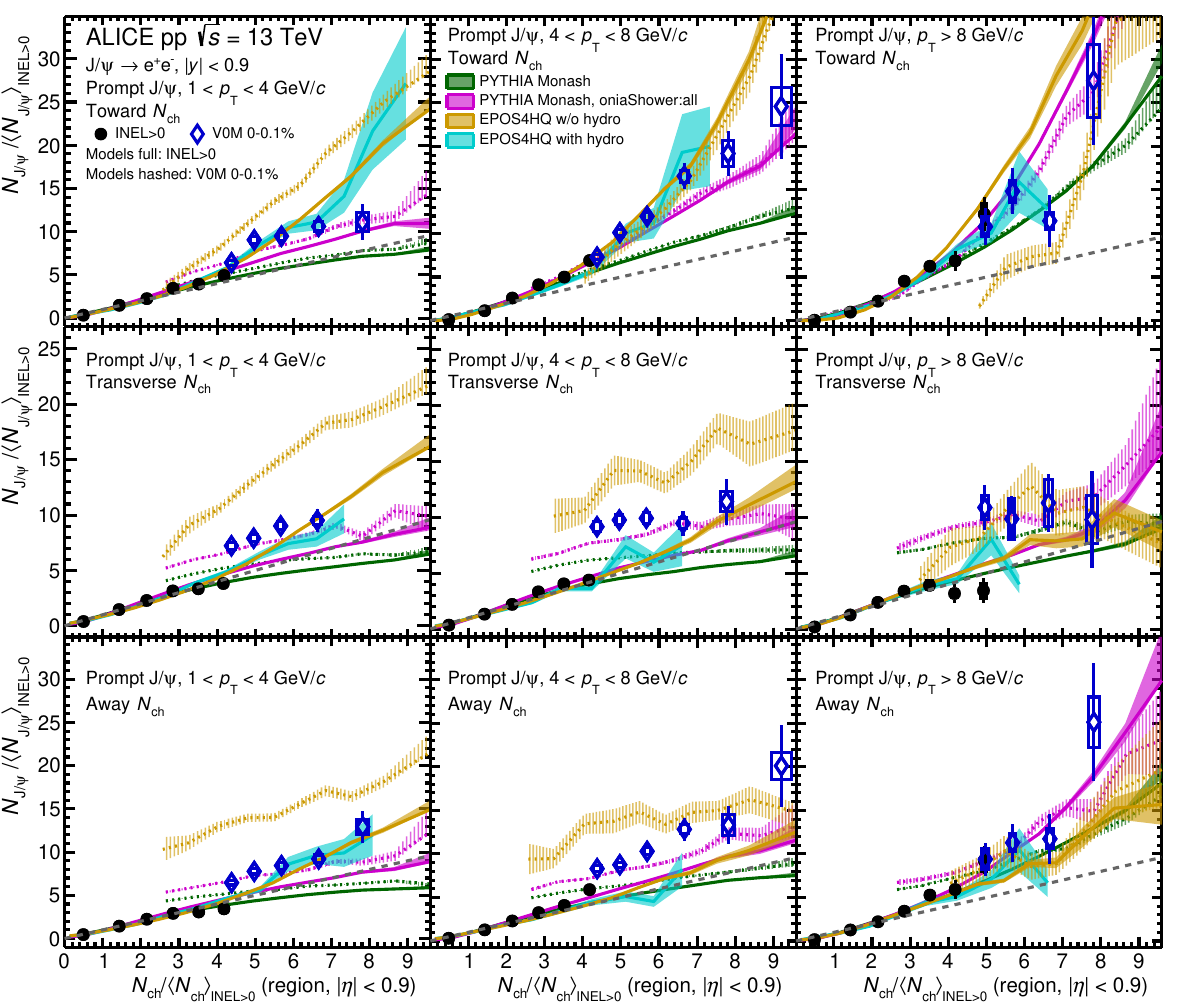}
  \caption{Self-normalized prompt \jpsi yields in the \pt intervals $1<\pt<4$~\GeVc (left column), $4<\pt<8$~\GeVc (middle column), and $\pt>8$~\GeVc (right column) as a function of the self-normalized charged-particle multiplicity in $|\eta|<0.9$ measured in toward (top row), transverse (middle row), and away (bottom row) azimuthal regions with respect to the \jpsi momentum direction. 
  The data are shown separately for the INEL$>$0 and V0M~0--0.1\% event classes and are compared to PYTHIA 8 simulations using different settings~\cite{Skands:2014pea, Cooke:2023ldt}, as well as EPOS4HQ with and without hydrodynamics activated~\cite{Werner:2023zvo, Zhao:2023ucp}. The dashed grey lines represent a linear increase with slope 1. The vertical bars and boxes indicate the statistical and systematic uncertainties, respectively. }
    \label{fig_results_prompt_regions_vspt}
\end{figure}

\begin{figure}[!htb]
  \centering
    \includegraphics[width=1.\linewidth]{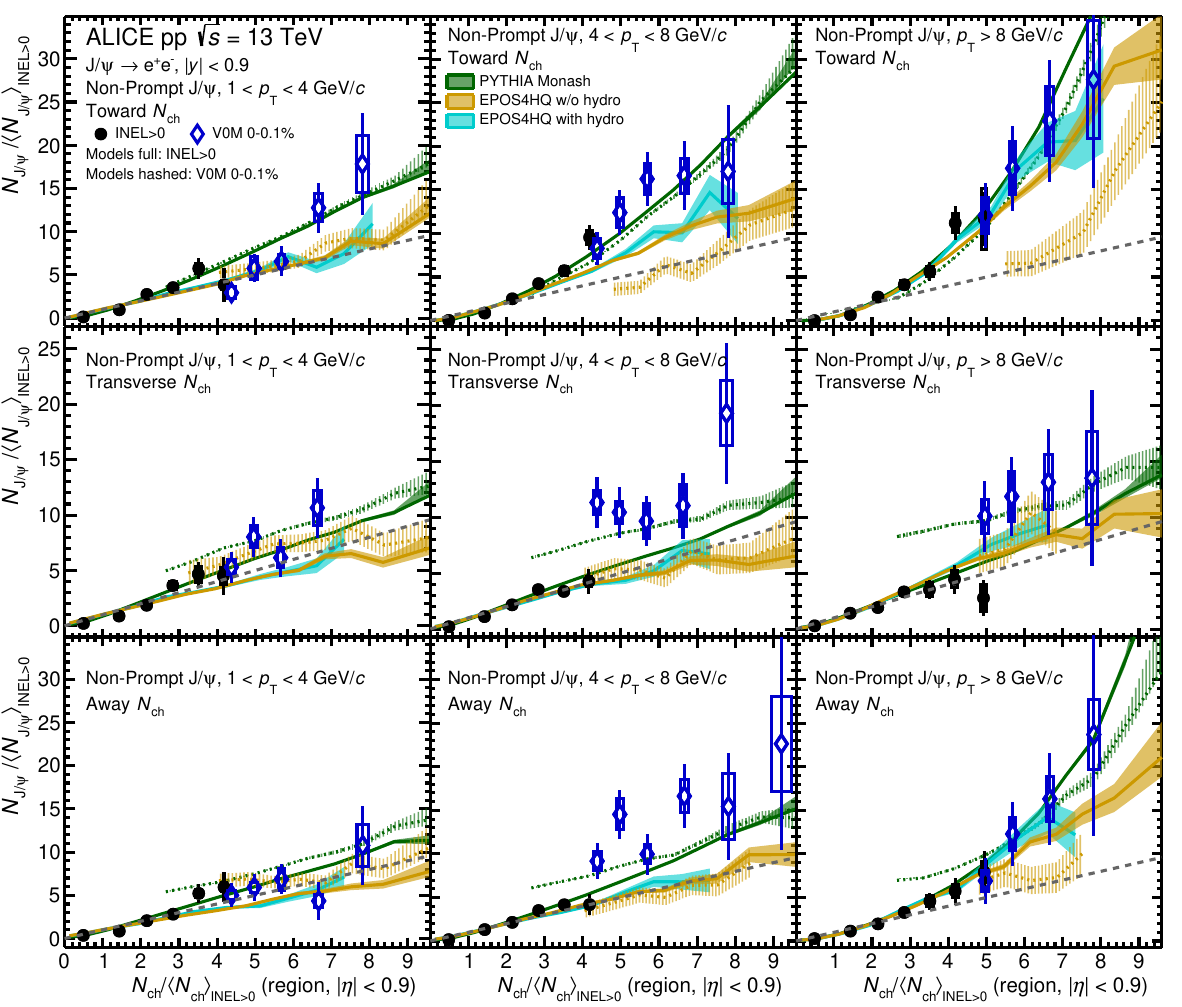}
  \caption{
  Self-normalized non-prompt \jpsi yields in the \pt intervals $1<\pt<4$~\GeVc (left column), $4<\pt<8$~\GeVc (middle column), and $\pt>8$~\GeVc (right column) as a function of the self-normalized charged-particle multiplicity in $|\eta|<0.9$ measured in toward (top row), transverse (middle row), and away (bottom row) azimuthal regions with respect to the \jpsi momentum direction. The data are shown separately for the INEL$>$0 and V0M~0--0.1\% event classes and are compared to PYTHIA 8 simulations using different settings~\cite{Skands:2014pea, Cooke:2023ldt}, as well as EPOS4HQ with and without hydrodynamics activated~\cite{Werner:2023zvo, Zhao:2023ucp}. The dashed grey lines represent a linear increase with slope 1. The vertical bars and boxes indicate the statistical and systematic uncertainties, respectively.
  }
    \label{fig_results_nonprompt_regions_vspt}
\end{figure}

Finally, Figures~\ref{fig_results_prompt_regions_vspt} (prompt) and~\ref{fig_results_nonprompt_regions_vspt} (non-prompt) show the \jpsi yields as a function of multiplicity in the three azimuthal regions, separately for the three \pt intervals. The strength of the dependence on the toward multiplicity for prompt \jpsi production grows fast with the \jpsi \pt, while only a moderate change is seen when using the transverse or away multiplicity. This suggests that the stronger increase of the multiplicity-dependent yields at higher \pt could come from higher associated particle production, e.g. from a harder jet, rather than from underlying event properties. The \pt dependence in the transverse and away region could also be affected by jet productions, such as the jets being more collimated or the more frequent occurrence of 3-jet topologies at high \pt. The additional presence of the \jpsi decay daughters at low \pt might also play a role. 
The opposite contributions  of these effects when varying \pt could hide a potential modification with increasing \pt of the underlying event in the transverse region.

For non-prompt \jpsi yields, a stronger increase at higher \pt is measured for both the toward and away regions. It may also be present for the transverse region, although with a smaller magnitude. This is presumably related to the production process, connected to the strong back-to-back correlation of the beauty pair from which the non-prompt \jpsi meson originates.

Similar to the data, PYTHIA and  EPOS4 calculations for prompt \jpsi production show a stronger increase of the correlation at higher \pt in the toward region, but not in the transverse region. In EPOS4, the non-prompt yields increase with multiplicity are stronger at higher \pt for all regions, which is not the case in PYTHIA. This could be due to the fact that the correlation between hard and soft scale is impacted by the saturation scale introduced in EPOS4. The effect of the saturation scale might influence all three regions as the saturation scale is a global event property.

\subsection{\texorpdfstring{\jpsi-to-D$^0$}{Jpsi-to-D0} ratio}

\begin{figure}[!htb]
  \centering
    \includegraphics[width=0.9\linewidth]{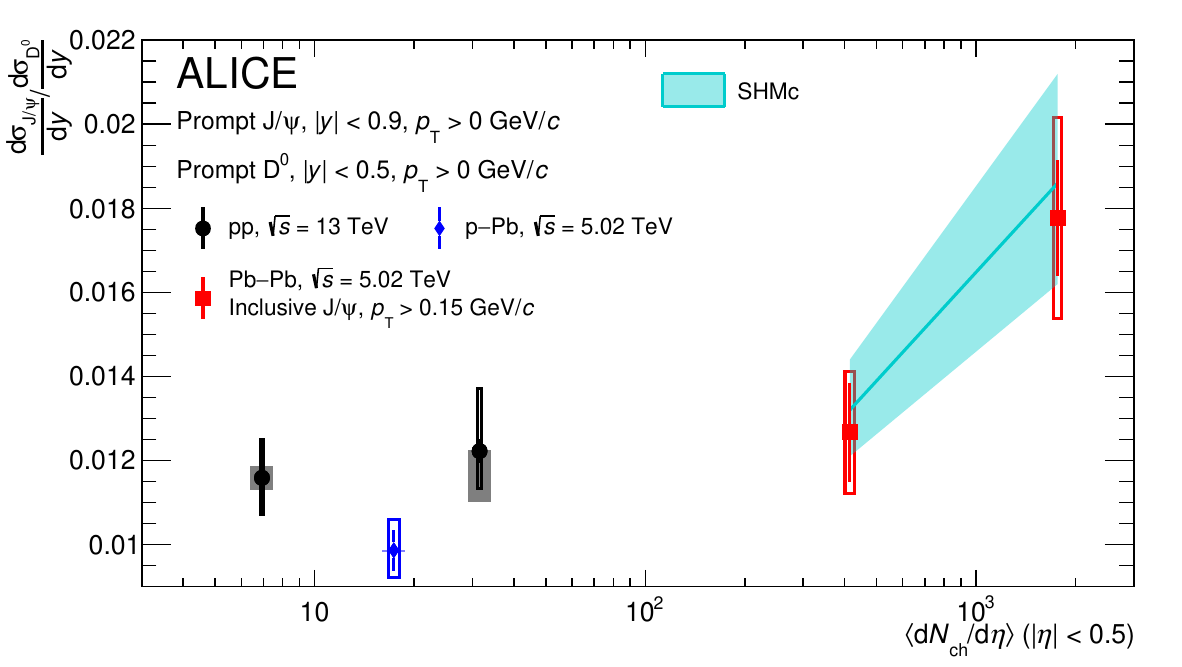}
  \caption{Ratio between the \pt-integrated prompt \jpsi and prompt D$^0$ yields as a function of charged-particle multiplicity within $|\eta|<0.5$ in several collision systems~\cite{ALICE:2019fhe,ALICE:2021lmn,ALICE:2023gco}. This is compared to the prediction from the Statistical Hadronization Model (SHMc)~\cite{Andronic:2021erx}. The vertical bars, empty and shaded boxes indicate the statistical, systematic and extrapolation uncertainties, respectively.
  }
    \label{fig_results_jpsioverD0}
\end{figure}

The ratio between prompt \jpsi and prompt D$^0$ yields is shown in Fig.~\ref{fig_results_jpsioverD0} as a function of charged-particle multiplicity across several colliding systems.  For pp collisions, the results are shown for integrated INEL$>$0 and V0M~0--0.1\% event classes. Data for prompt D$^0$ and for other collision systems are taken from previous ALICE measurements in pp collisions at \sqrts{13}~\cite{ALICE:2021npz}, p--Pb collisions at \sqrtsnn{5.02}~\cite{ALICE:2019fhe,ALICE:2021lmn}, and Pb--Pb collisions at the same energy~\cite{ALICE:2023gco}. In the latter case, inclusive \jpsi mesons are considered, and a momentum selection $\pt>0.15$ GeV/$c$ removes most of the contribution from photoproduction. The corresponding multiplicity values are also taken from previous ALICE measurements~\cite{ALICE:2020swj,ALICE:2012xs,ALICE:2015juo}.

The prompt \jpsi yield is extrapolated down to $\pt=0$~\GeVc. In order to do so, the inclusive yield is extracted for $\pt <1$~\GeVc, for both INEL$>$0 and V0M~0--0.1\% classes, while the non-prompt yield for $\pt <1$~\GeVc is subtracted from it. The non-prompt yield is estimated using PYTHIA with CR mode 2, scaled in order to reproduce the non-prompt yield in the data for $\pt>1$~\GeVc. The prompt yield with $\pt<1$~\GeVc is then summed to the one with $\pt>1$~\GeVc. An extrapolation uncertainty is estimated by changing the model used for the extrapolation. The variations tested are: PYTHIA with Monash tune and CR mode 3, FONLL calculations with central value, upward and downward variations of the theory uncertainties. The maximum deviation considering all variations is assigned as extrapolation uncertainty in the INEL$>$0 case. In the V0M~0--0.1\% case, because FONLL is not available, the envelope of all the PYTHIA variations is added in quadrature to the FONLL uncertainty for INEL$>$0. An uncertainty due to the scaling of the models for the non-prompt yield is also assigned. This is done by moving the data points used for the scaling by the sum in quadrature of statistical and systematic uncertainties. The tracking, MB trigger and luminosity uncertainties are assumed to cancel in the ratio.

In pp collisions, no modification of the ratio can be observed within uncertainties. The ratio in semicentral Pb--Pb collisions is compatible with the one in pp collisions within uncertainties, while the data suggest an increase for central collisions. The Pb--Pb data are described very well by the SHMc\,\cite{Andronic:2021erx}, which assumes that the \jpsi and D$^0$ mesons are produced at the QCD crossover phase boundary. While the increase in central Pb--Pb collisions could be due to regeneration of the \jpsi from uncorrelated c and $\overline{\rm c}$ quarks, the current data, although they have large uncertainties, do not allow to conclude on such a regeneration in pp collisions at high multiplicity.

%% file: E_Summary.tex

The evolution of the yields of prompt and non-prompt \jpsi mesons with charged-particle multiplicity has been measured in pp collisions at \sqrts{13}. The charged-particle multiplicity is measured within $|\eta|<0.9$, and separated in three azimuthal regions.

Both prompt and non-prompt \jpsi self-normalized yields show a similar stronger-than-linear increase with self-normalized multiplicity. A hint of a slightly stronger increase for non-prompt \jpsi yields is revealed by the evolution of the non-prompt fraction with multiplicity.
The multiplicity dependence of the \jpsi production is stronger at higher \pt and when the multiplicity is evaluated in an azimuthal range towards the direction of the \jpsi meson. In addition, the ratio between prompt \jpsi and prompt D$^0$ yields does not show significant difference, within large uncertainties, between INEL$>$0 pp, high-multiplicity pp and semicentral Pb--Pb collisions.
However, the measurements with the high-multiplicity (HM) trigger used to collect data at very high forward multiplicity show a significant bias in some cases. The results can nevertheless be interpreted by comparison with PYTHIA where a selection reproducing the effect of the trigger has been applied.
PYTHIA calculations with several settings reproduce the results for non-prompt \jpsi production. In contrast, for prompt \jpsi production, the data are consistently underestimated by the Monash tune. The oniaShower setting is necessary to reproduce the measurement. CGC-based models are unable to reproduce the correlations in all the \pt intervals, while EPOS4 calculations overestimate the prompt and underestimate the non-prompt \jpsi results.

These measurements underline that autocorrelations with particles coming from the same production process as the \jpsi meson are playing a role in the correlation between \jpsi yields and multiplicity. Their role is significant  for explaining the stronger increase of the multiplicity-dependent yields at higher \pt. The correlation remains also strong for the azimuthal region transverse to the \jpsi direction. Therefore, it can be concluded that, in addition to the autocorrelations, there is also an effect on the correlation caused by the scales at play (relatively-hard for the \jpsi meson, soft for the charged-particle multiplicity) in all regions.

During the Run 3 of the LHC, a larger dataset of pp collisions is being collected by ALICE without hardware triggers and with improved vertex pointing resolution. This will allow for more precise and less trigger-biased measurements of the multiplicity dependence of prompt and non-prompt \jpsi production to be conducted in the future.

%% file: fa_2026-02-26_Opt_C.tex

The ALICE Collaboration would like to thank all its engineers and technicians for their invaluable contributions to the construction of the experiment and the CERN accelerator teams for the outstanding performance of the LHC complex.
The ALICE Collaboration gratefully acknowledges the resources and support provided by all Grid centres and the Worldwide LHC Computing Grid (WLCG) collaboration.
The ALICE Collaboration acknowledges the following funding agencies for their support in building and running the ALICE detector:
A. I. Alikhanyan National Science Laboratory (Yerevan Physics Institute) Foundation (ANSL), State Committee of Science and World Federation of Scientists (WFS), Armenia;
Austrian Academy of Sciences, Austrian Science Fund (FWF): [M 2467-N36] and Nationalstiftung f\"{u}r Forschung, Technologie und Entwicklung, Austria;
Ministry of Communications and High Technologies, National Nuclear Research Center, Azerbaijan;
Rede Nacional de Física de Altas Energias (Renafae), Financiadora de Estudos e Projetos (Finep), Funda\c{c}\~{a}o de Amparo \`{a} Pesquisa do Estado de S\~{a}o Paulo (FAPESP) and The Sao Paulo Research Foundation  (FAPESP), Brazil;
Bulgarian Ministry of Education and Science, within the National Roadmap for Research Infrastructures 2020-2027 (object CERN), Bulgaria;
Ministry of Education of China (MOEC) , Ministry of Science \& Technology of China (MSTC) and National Natural Science Foundation of China (NSFC), China;
Ministry of Science and Education and Croatian Science Foundation, Croatia;
Centro de Aplicaciones Tecnol\'{o}gicas y Desarrollo Nuclear (CEADEN), Cubaenerg\'{\i}a, Cuba;
Ministry of Education, Youth and Sports of the Czech Republic, Czech Republic;
The Danish Council for Independent Research | Natural Sciences, the VILLUM FONDEN and Danish National Research Foundation (DNRF), Denmark;
Helsinki Institute of Physics (HIP), Finland;
Commissariat \`{a} l'Energie Atomique (CEA) and Institut National de Physique Nucl\'{e}aire et de Physique des Particules (IN2P3) and Centre National de la Recherche Scientifique (CNRS), France;
Bundesministerium f\"{u}r Forschung, Technologie und Raumfahrt (BMFTR) and GSI Helmholtzzentrum f\"{u}r Schwerionenforschung GmbH, Germany;
National Research, Development and Innovation Office, Hungary;
Department of Atomic Energy Government of India (DAE), Department of Science and Technology, Government of India (DST), University Grants Commission, Government of India (UGC) and Council of Scientific and Industrial Research (CSIR), India;
National Research and Innovation Agency - BRIN, Indonesia;
Istituto Nazionale di Fisica Nucleare (INFN), Italy;
Japanese Ministry of Education, Culture, Sports, Science and Technology (MEXT) and Japan Society for the Promotion of Science (JSPS) KAKENHI, Japan;
Consejo Nacional de Ciencia (CONACYT) y Tecnolog\'{i}a, through Fondo de Cooperaci\'{o}n Internacional en Ciencia y Tecnolog\'{i}a (FONCICYT) and Direcci\'{o}n General de Asuntos del Personal Academico (DGAPA), Mexico;
Nederlandse Organisatie voor Wetenschappelijk Onderzoek (NWO), Netherlands;
The Research Council of Norway, Norway;
Pontificia Universidad Cat\'{o}lica del Per\'{u}, Peru;
Ministry of Science and Higher Education, National Science Centre and WUT ID-UB, Poland;
Korea Institute of Science and Technology Information and National Research Foundation of Korea (NRF), Republic of Korea;
Ministry of Education and Scientific Research, Institute of Atomic Physics, Ministry of Research and Innovation and Institute of Atomic Physics and Universitatea Nationala de Stiinta si Tehnologie Politehnica Bucuresti, Romania;
Ministerstvo skolstva, vyskumu, vyvoja a mladeze SR, Slovakia;
National Research Foundation of South Africa, South Africa;
Swedish Research Council (VR) and Knut \& Alice Wallenberg Foundation (KAW), Sweden;
European Organization for Nuclear Research, Switzerland;
Suranaree University of Technology (SUT), National Science and Technology Development Agency (NSTDA) and National Science, Research and Innovation Fund (NSRF via PMU-B B05F650021), Thailand;
Turkish Energy, Nuclear and Mineral Research Agency (TENMAK), Turkey;
National Academy of  Sciences of Ukraine, Ukraine;
Science and Technology Facilities Council (STFC), United Kingdom;
National Science Foundation of the United States of America (NSF) and United States Department of Energy, Office of Nuclear Physics (DOE NP), United States of America.
In addition, individual groups or members have received support from:
Czech Science Foundation (grant no. 23-07499S), Czech Republic;
FORTE project, reg.\ no.\ CZ.02.01.01/00/22\_008/0004632, Czech Republic, co-funded by the European Union, Czech Republic;
European Research Council (grant nos. 101220549, 950692), European Union;
Deutsche Forschungs Gemeinschaft (DFG, German Research Foundation) ``Neutrinos and Dark Matter in Astro- and Particle Physics'' (grant no. SFB 1258), Germany;
CONVECS project, CUP C97H23001700002 FESR 2021-2027 program, Italy;
FAIR - Future Artificial Intelligence Research, funded by the NextGenerationEU program (Italy).

%% file: AppendixA_DaughtersRemoval.tex
\begin{figure}[htb]
  \centering
    \includegraphics[width=0.9\linewidth]{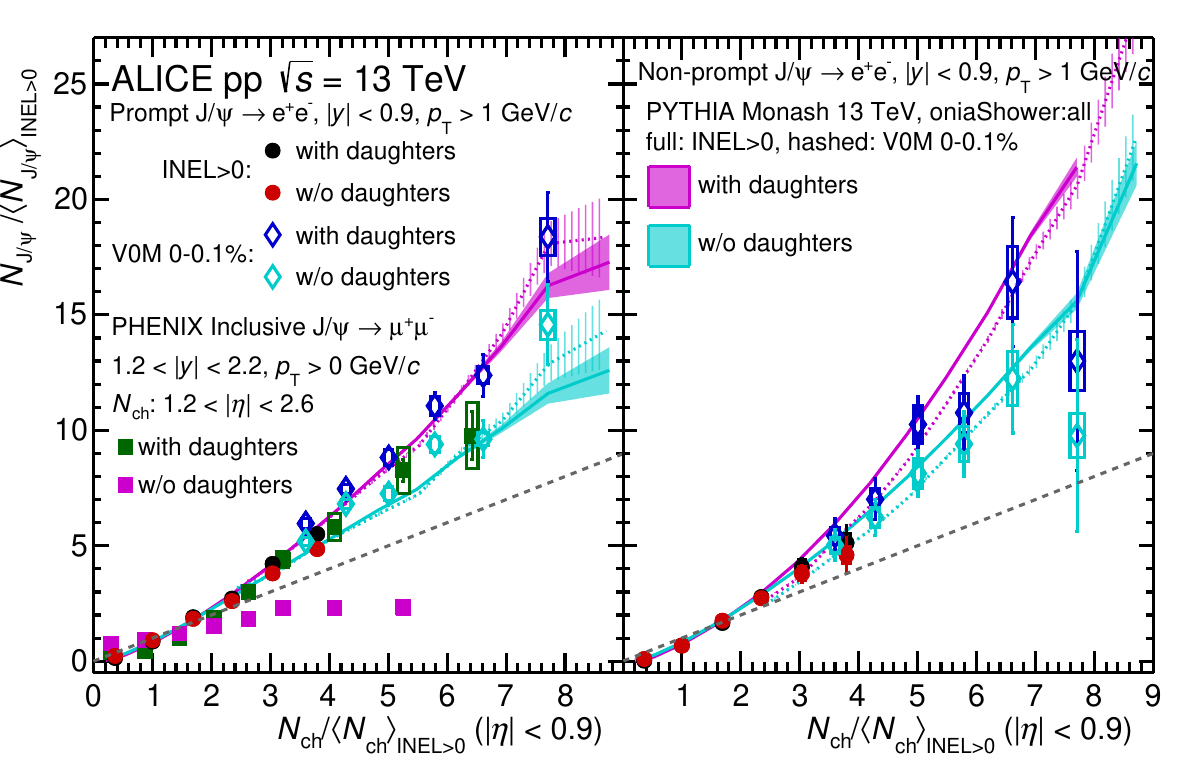}
  \caption{Self-normalized prompt (left) and non-prompt (right) \jpsi yields with $\pt>1$~\GeVc as a function of self-normalized charged-particle multiplicity within $|\eta|<0.9$, in the INEL$>$0 (full markers) and V0M 0--0.1\% (open markers) event classes. The case when the \jpsi decay daughters are included in the multiplicity calculation is compared with the case when both \jpsi decay daughters are removed from the multiplicity calculation. The yields are compared to PYTHIA 8~\cite{Bierlich:2022pfr} using oniaShower:all settings. The dashed grey lines represent a linear increase with slope 1. The vertical bars and boxes indicate the statistical and systematic uncertainties, respectively. The prompt \jpsi yields are also compared to the inclusive \jpsi yields from the PHENIX collaboration~\cite{PHENIX:2024dqs}.}
    \label{fig_results_removelegs}
\end{figure}

When \jpsi production and multiplicity are measured in the same rapidity region, the decay products of the \jpsi meson may enter in the multiplicity count, potentially introducing a bias.
In Fig.~\ref{fig_results_removelegs}, the self-normalized yields of prompt and non-prompt \jpsi mesons with $\pt > 1$~\GeVc are shown as a function of multiplicity, with and without including the \jpsi decay daughters in the multiplicity calculation. Due to different multiplicity distributions at midrapidity, removing these two decay daughters has a different impact on INEL$>$0 and HM-triggered events. Therefore, for this figure only, 
when removing the \jpsi decay daughters from the multiplicity, a multiplicity-dependent weight 
is applied to \jpsi candidates in the HM sample. This weight is defined as

\begin{equation}
    \frac{N^{\rm HM}_{\rm evt}\left(N_{\text{trks}}^{\text{w/o daughters}}\right)}{N^{\rm INEL>0}_{\rm evt}\left(N_{\text{trks}}^{\text{w/o daughters}}\right)} \frac{N^{\rm INEL>0}_{\rm evt}(N_{\text{trks}})}{N^{\rm HM}_{\rm evt}(N_{\text{trks}})}\, ,
\end{equation}

where $N_{\rm evt}$ represents the event multiplicity distributions, evaluated at the value of $N_{\rm trks}$ in the event containing the \jpsi meson, either with or without excluding the \jpsi decay daughters from $N_{\rm trks}$ calculation. This weight ensures that the HM trigger bias is similar when including and removing the \jpsi decay daughters.

The decrease of the self-normalized yields when \jpsi decay daughters are removed is found to be similar between prompt and non-prompt \jpsi production. 
It is reproduced in PYTHIA, and is explained by the fact that the events with \jpsi mesons are associated with a lower multiplicity value when the \jpsi decay daughters are removed compared to when they are included. The total number of unbiased events with this lower multiplicity value is higher, thus reducing the \jpsi yield per event. This shows the presence of autocorrelations, in this case brought about by the \jpsi decay daughters. The results are also compared to the PHENIX measurement at $\sqrt{s}=200$ GeV~\cite{PHENIX:2024dqs}. The impact of removing the \jpsi decay daughters from the multiplicity on these results was found to be very significant at RHIC. However, at LHC energies, the observed effect is much smaller.  This is explained by a higher average charged-particle multiplicity and a less steep decrease of the multiplicity distribution at LHC energies than at RHIC energies. The driving effects on the multiplicity distribution are the average number of MPIs and its fluctuations, which increase significantly from RHIC to LHC energies.

The multiplicity-dependent inclusive \jpsi yield is very similar when the multiplicity is estimated directly at midrapidity and when it is selected at forward rapidity and converted to midrapidity multiplicity values~\cite{ALICE:2020msa}. The V0 estimator does not contain the \jpsi decay daughters, but it could contain additional multiplicity indissociable from the \jpsi production, such as a recoil jet.
Thus, the bias present when estimating the multiplicity at midrapidity and including the \jpsi decay daughters in the estimation could be small. 
In contrast, when removing the \jpsi decay daughters from the multiplicity calculation, the additional multiplicity brought along with the presence of the \jpsi meson is not accounted for. In a hypothetical baseline case for which hard and soft particle production would not differ, the measurement of \jpsi production as a function of the total multiplicity would be expected to be linear. However, measuring \jpsi yields only as a function of the multiplicity uncorrelated to itself would bias the measurement towards a weaker increase, and the baseline would not be linear.

Thus, all the measurements from Sec.~\ref{sec:meas}, as well as the calculations from MC generators, are shown with inclusion of the \jpsi decay daughters in the multiplicity estimation. In models which are not MC generators, the \jpsi decay daughters are not included or removed explicitly, and the impact of the absence of a clear treatment of these autocorrelations is not completely known.

%% file: AppendixB_AdditionalModels.tex
\begin{figure}[!htb]
  \centering
    \includegraphics[width=0.8\linewidth]{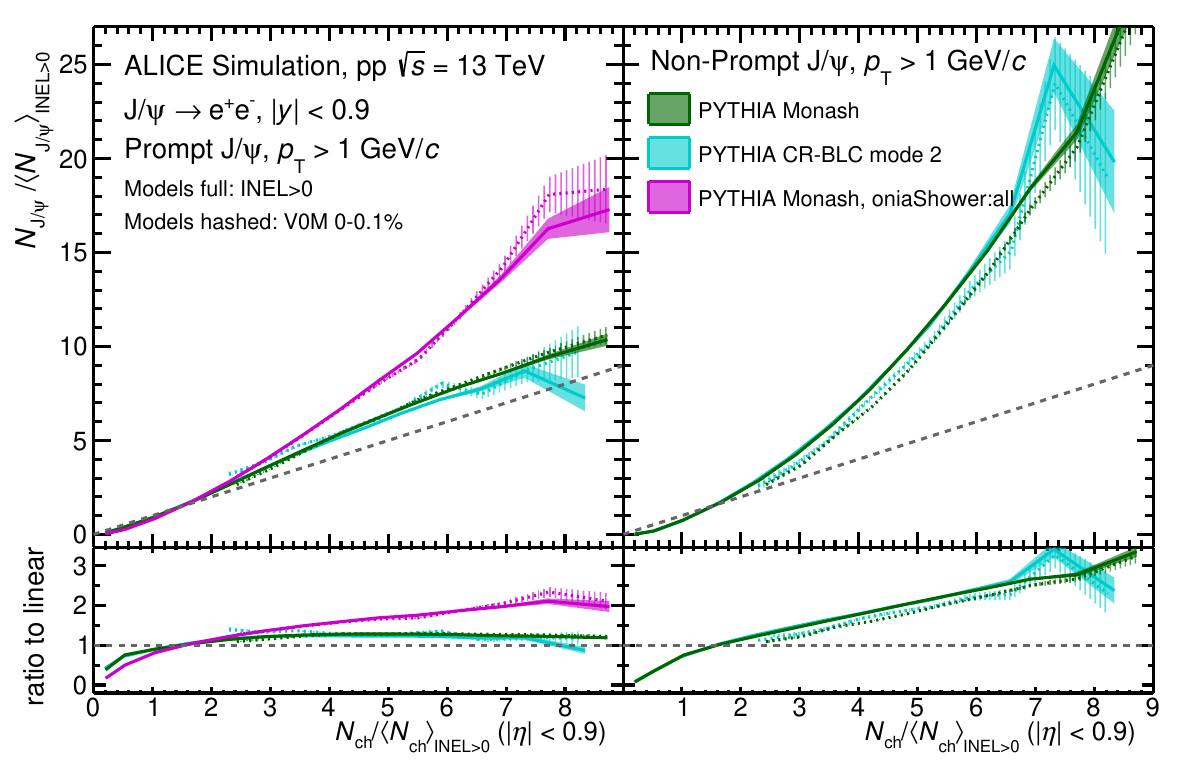}
  \caption{Self-normalized prompt (left) and non-prompt (right) \jpsi yields with $\pt>1$~\GeVc as a function of self-normalized charged-particle multiplicity at midrapidity from PYTHIA 8 simulations with different settings~\cite{Skands:2014pea, Cooke:2023ldt, Christiansen:2015yqa}. The yields are shown separately for the INEL$>$0 (full) and V0M~0-0.1\% (hashed) event classes. Lower panels show the ratios to the expectation from a linear increase.}
\label{fig_results_prompt_nonprompt_pythia}
\end{figure}

\begin{figure}[!htb]
  \centering
    \includegraphics[width=1.\linewidth]{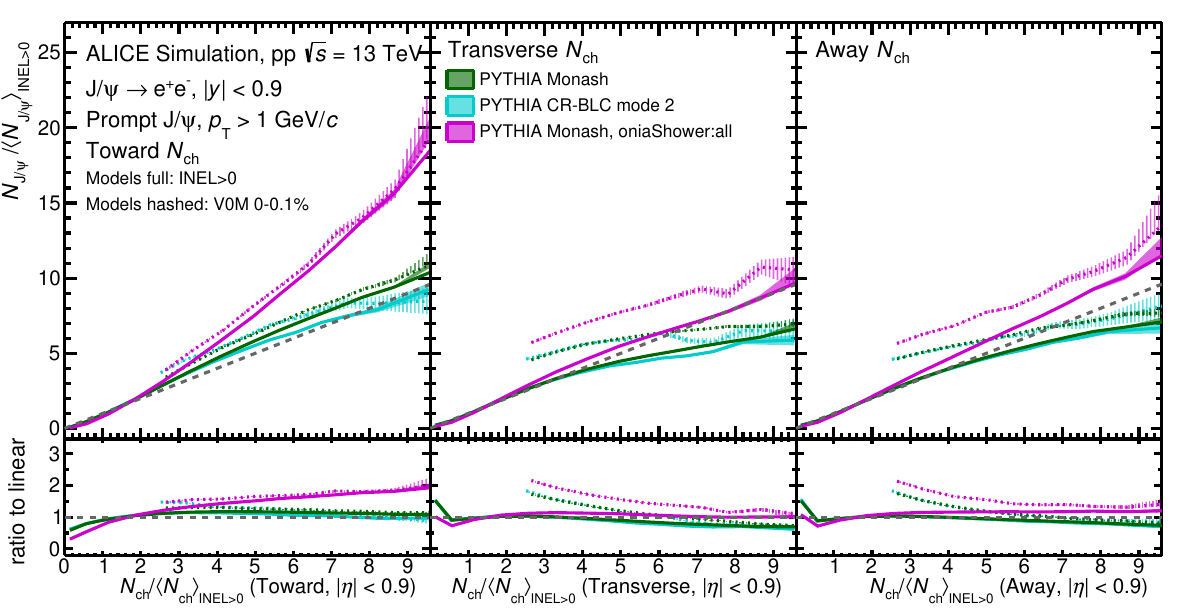}
  \caption{Self-normalized prompt \jpsi yields with $\pt>1$~\GeVc, taken from PYTHIA 8 simulations with different settings~\cite{Skands:2014pea, Cooke:2023ldt, Christiansen:2015yqa}, as a function of the self-normalized charged-particle multiplicity in $|\eta|<0.9$  in the toward (left), transverse (middle), and away (right) azimuthal regions with respect to the \jpsi momentum direction. 
  The yields are shown separately for the INEL$>$0 (full) and V0M~0-0.1\% (hashed) event classes. Lower panels show the ratios to the expectation from a linear increase.}
    \label{fig_results_prompt_regions_pythia}
\end{figure}

Figure~\ref{fig_results_prompt_nonprompt_pythia} shows the predictions from different PYTHIA settings (Monash~\cite{Skands:2014pea}, CR-BLC mode 2~\cite{Christiansen:2015yqa}, and oniaShower~\cite{Cooke:2023ldt}) for prompt (left) and non-prompt (right) self-normalized yields. Fig.~\ref{fig_results_prompt_regions_pythia} shows the PYTHIA comparison for prompt \jpsi as a function of multiplicity in azimuthal regions. For all these predictions, CR-BLC mode 2 yields are found to be very close to Monash, while the oniaShower settings give significantly larger yields for prompt \jpsi.

%% file: Alice_Authorlist_2026-02-26_Opt_C.tex
\begin{flushleft} 
\small

D.A.H.~Abdallah\,\orcidlink{0000-0003-4768-2718}\,$^{\rm 134}$, 
I.J.~Abualrob\,\orcidlink{0009-0005-3519-5631}\,$^{\rm 112}$, 
S.~Acharya\,\orcidlink{0000-0002-9213-5329}\,$^{\rm 49}$, 
K.~Agarwal\,\orcidlink{0000-0001-5781-3393}\,$^{\rm II,}$$^{\rm 23}$, 
G.~Aglieri Rinella\,\orcidlink{0000-0002-9611-3696}\,$^{\rm 32}$, 
L.~Aglietta\,\orcidlink{0009-0003-0763-6802}\,$^{\rm 24}$, 
N.~Agrawal\,\orcidlink{0000-0003-0348-9836}\,$^{\rm 25}$, 
Z.~Ahammed\,\orcidlink{0000-0001-5241-7412}\,$^{\rm 132}$, 
S.~Ahmad\,\orcidlink{0000-0003-0497-5705}\,$^{\rm 15}$, 
I.~Ahuja\,\orcidlink{0000-0002-4417-1392}\,$^{\rm 36}$
% ...

\section*{Affiliation Notes}

$^{\rm I}$ Deceased\\
$^{\rm II}$ Also at: INFN Trieste, Trieste, Italy\\
$^{\rm III}$ Also at: Fondazione Bruno Kessler (FBK), Trento, Italy\\
$^{\rm IV}$ Also at: Max-Planck-Institut fur Physik, Munich, Germany\\
$^{\rm V}$ Also at: Czech Technical University in Prague, Prague, Czech Republic\\
$^{\rm VI}$ Also at: Instituto de Fisica da Universidade de Sao Paulo\\
$^{\rm VII}$ Also at: Dipartimento DET del Politecnico di Torino, Turin, Italy\\
$^{\rm VIII}$ Also at: University College of Dublin, Dublin, Ireland\\
$^{\rm IX}$ Also at: Department of Applied Physics, Aligarh Muslim University, Aligarh, India\\
$^{\rm X}$ Also at: Institute of Theoretical Physics, University of Wroclaw, Wroclaw, Poland\\
$^{\rm XI}$ Also at: Facultad de Ciencias, Universidad Nacional Aut\'{o}noma de M\'{e}xico, Mexico City, Mexico\\

\section*{Collaboration Institutes}

$^{1}$ A.I. Alikhanyan National Science Laboratory (Yerevan Physics Institute) Foundation, Yerevan, Armenia\\
$^{2}$ AGH University of Krakow, Cracow, Poland\\
$^{3}$ Bogolyubov Institute for Theoretical Physics, National Academy of Sciences of Ukraine, Kyiv, Ukraine\\
$^{4}$ Bose Institute, Department of Physics  and Centre for Astroparticle Physics and Space Science (CAPSS), Kolkata, India\\
$^{5}$ California Polytechnic State University, San Luis Obispo, California, United States\\
$^{6}$ Central China Normal University, Wuhan, China\\
$^{7}$ Centro de Aplicaciones Tecnol\'{o}gicas y Desarrollo Nuclear (CEADEN), Havana, Cuba\\
$^{8}$ Centro de Investigaci\'{o}n y de Estudios Avanzados (CINVESTAV), Mexico City and M\'{e}rida, Mexico\\
$^{9}$ Chicago State University, Chicago, Illinois, United States\\
$^{10}$ China Nuclear Data Center, China Institute of Atomic Energy, Beijing, China\\
$^{11}$ China University of Geosciences, Wuhan, China\\
$^{12}$ Chungbuk National University, Cheongju, Republic of Korea\\
$^{13}$ Comenius University Bratislava, Faculty of Mathematics, Physics and Informatics, Bratislava, Slovak Republic\\
$^{14}$ Creighton University, Omaha, Nebraska, United States\\
$^{15}$ Department of Physics, Aligarh Muslim University, Aligarh, India\\
$^{16}$ Department of Physics, Pusan National University, Pusan, Republic of Korea\\
$^{17}$ Department of Physics, Sejong University, Seoul, Republic of Korea\\
$^{18}$ Department of Physics, University of California, Berkeley, California, United States\\
$^{19}$ Department of Physics, University of Oslo, Oslo, Norway\\
$^{20}$ Department of Physics and Technology, University of Bergen, Bergen, Norway\\
$^{21}$ Dipartimento di Fisica, Universit\`{a} di Pavia, Pavia, Italy\\
$^{22}$ Dipartimento di Fisica dell'Universit\`{a} and Sezione INFN, Cagliari, Italy\\
$^{23}$ Dipartimento di Fisica dell'Universit\`{a} and Sezione INFN, Trieste, Italy\\
$^{24}$ Dipartimento di Fisica dell'Universit\`{a} and Sezione INFN, Turin, Italy\\
$^{25}$ Dipartimento di Fisica e Astronomia dell'Universit\`{a} and Sezione INFN, Bologna, Italy\\
$^{26}$ Dipartimento di Fisica e Astronomia dell'Universit\`{a} and Sezione INFN, Catania, Italy\\
$^{27}$ Dipartimento di Fisica e Astronomia dell'Universit\`{a} and Sezione INFN, Padova, Italy\\
$^{28}$ Dipartimento di Fisica `E.R.~Caianiello' dell'Universit\`{a} and Gruppo Collegato INFN, Salerno, Italy\\
$^{29}$ Dipartimento DISAT del Politecnico and Sezione INFN, Turin, Italy\\
$^{30}$ Dipartimento di Scienze MIFT, Universit\`{a} di Messina, Messina, Italy\\
$^{31}$ Dipartimento Interateneo di Fisica `M.~Merlin' and Sezione INFN, Bari, Italy\\
$^{32}$ European Organization for Nuclear Research (CERN), Geneva, Switzerland\\
$^{33}$ Faculty of Electrical Engineering, Mechanical Engineering and Naval Architecture, University of Split, Split, Croatia\\
$^{34}$ Faculty of Nuclear Sciences and Physical Engineering, Czech Technical University in Prague, Prague, Czech Republic\\
$^{35}$ Faculty of Physics, Sofia University, Sofia, Bulgaria\\
$^{36}$ Faculty of Science, P.J.~\v{S}af\'{a}rik University, Ko\v{s}ice, Slovak Republic\\
$^{37}$ Faculty of Technology, Environmental and Social Sciences, Bergen, Norway\\
$^{38}$ Frankfurt Institute for Advanced Studies, Johann Wolfgang Goethe-Universit\"{a}t Frankfurt, Frankfurt, Germany\\
$^{39}$ Fudan University, Shanghai, China\\
$^{40}$ Gauhati University, Department of Physics, Guwahati, India\\
$^{41}$ Helmholtz-Institut f\"{u}r Strahlen- und Kernphysik, Rheinische Friedrich-Wilhelms-Universit\"{a}t Bonn, Bonn, Germany\\
$^{42}$ Helsinki Institute of Physics (HIP), Helsinki, Finland\\
$^{43}$ High Energy Physics Group,  Universidad Aut\'{o}noma de Puebla, Puebla, Mexico\\
$^{44}$ Horia Hulubei National Institute of Physics and Nuclear Engineering, Bucharest, Romania\\
$^{45}$ HUN-REN Wigner Research Centre for Physics, Budapest, Hungary\\
$^{46}$ Indian Institute of Technology Bombay (IIT), Mumbai, India\\
$^{47}$ Indian Institute of Technology Indore, Indore, India\\
$^{48}$ INFN, Laboratori Nazionali di Frascati, Frascati, Italy\\
$^{49}$ INFN, Sezione di Bari, Bari, Italy\\
$^{50}$ INFN, Sezione di Bologna, Bologna, Italy\\
$^{51}$ INFN, Sezione di Cagliari, Cagliari, Italy\\
$^{52}$ INFN, Sezione di Catania, Catania, Italy\\
$^{53}$ INFN, Sezione di Padova, Padova, Italy\\
$^{54}$ INFN, Sezione di Pavia, Pavia, Italy\\
$^{55}$ INFN, Sezione di Torino, Turin, Italy\\
$^{56}$ INFN, Sezione di Trieste, Trieste, Italy\\
$^{57}$ Inha University, Incheon, Republic of Korea\\
$^{58}$ Institute for Gravitational and Subatomic Physics (GRASP), Utrecht University/Nikhef, Utrecht, Netherlands\\
$^{59}$ Institute of Experimental Physics, Slovak Academy of Sciences, Ko\v{s}ice, Slovak Republic\\
$^{60}$ Institute of Physics, Homi Bhabha National Institute, Bhubaneswar, India\\
$^{61}$ Institute of Physics of the Czech Academy of Sciences, Prague, Czech Republic\\
$^{62}$ Institute of Space Science (ISS), Bucharest, Romania\\
$^{63}$ Institut f\"{u}r Kernphysik, Johann Wolfgang Goethe-Universit\"{a}t Frankfurt, Frankfurt, Germany\\
$^{64}$ Instituto de Ciencias Nucleares, Universidad Nacional Aut\'{o}noma de M\'{e}xico, Mexico City, Mexico\\
$^{65}$ Instituto de F\'{i}sica, Universidade Federal do Rio Grande do Sul (UFRGS), Porto Alegre, Brazil\\
$^{66}$ Instituto de F\'{\i}sica, Universidad Nacional Aut\'{o}noma de M\'{e}xico, Mexico City, Mexico\\
$^{67}$ iThemba LABS, National Research Foundation, Somerset West, South Africa\\
$^{68}$ Jeonbuk National University, Jeonju, Republic of Korea\\
$^{69}$ Korea Institute of Science and Technology Information, Daejeon, Republic of Korea\\
$^{70}$ Laboratoire de Physique Subatomique et de Cosmologie, Universit\'{e} Grenoble-Alpes, CNRS-IN2P3, Grenoble, France\\
$^{71}$ Lawrence Berkeley National Laboratory, Berkeley, California, United States\\
$^{72}$ Lund University Department of Physics, Division of Particle Physics, Lund, Sweden\\
$^{73}$ Marietta Blau Institute, Vienna, Austria\\
$^{74}$ Nagasaki Institute of Applied Science, Nagasaki, Japan\\
$^{75}$ Nara Women{'}s University (NWU), Nara, Japan\\
$^{76}$ National Centre for Nuclear Research, Warsaw, Poland\\
$^{77}$ National Institute of Science Education and Research, Homi Bhabha National Institute, Jatni, India\\
$^{78}$ National Nuclear Research Center, Baku, Azerbaijan\\
$^{79}$ National Research and Innovation Agency - BRIN, Jakarta, Indonesia\\
$^{80}$ Niels Bohr Institute, University of Copenhagen, Copenhagen, Denmark\\
$^{81}$ Nikhef, National institute for subatomic physics, Amsterdam, Netherlands\\
$^{82}$ Nuclear Physics Group, STFC Daresbury Laboratory, Daresbury, United Kingdom\\
$^{83}$ Nuclear Physics Institute of the Czech Academy of Sciences, Husinec-\v{R}e\v{z}, Czech Republic\\
$^{84}$ Oak Ridge National Laboratory, Oak Ridge, Tennessee, United States\\
$^{85}$ Ohio State University, Columbus, Ohio, United States\\
$^{86}$ Physics department, Faculty of science, University of Zagreb, Zagreb, Croatia\\
$^{87}$ Physics Department, Panjab University, Chandigarh, India\\
$^{88}$ Physics Department, University of Jammu, Jammu, India\\
$^{89}$ Physics Program and International Institute for Sustainability with Knotted Chiral Meta Matter (WPI-SKCM$^{2}$), Hiroshima University, Hiroshima, Japan\\
$^{90}$ Physikalisches Institut, Eberhard-Karls-Universit\"{a}t T\"{u}bingen, T\"{u}bingen, Germany\\
$^{91}$ Physikalisches Institut, Ruprecht-Karls-Universit\"{a}t Heidelberg, Heidelberg, Germany\\
$^{92}$ Physik Department, Technische Universit\"{a}t M\"{u}nchen, Munich, Germany\\
$^{93}$ Politecnico di Bari and Sezione INFN, Bari, Italy\\
$^{94}$ Research Division and ExtreMe Matter Institute EMMI, GSI Helmholtzzentrum f\"ur Schwerionenforschung GmbH, Darmstadt, Germany\\
$^{95}$ Saga University, Saga, Japan\\
$^{96}$ Saha Institute of Nuclear Physics, Homi Bhabha National Institute, Kolkata, India\\
$^{97}$ School of Physics and Astronomy, University of Birmingham, Birmingham, United Kingdom\\
$^{98}$ Secci\'{o}n F\'{\i}sica, Departamento de Ciencias, Pontificia Universidad Cat\'{o}lica del Per\'{u}, Lima, Peru\\
$^{99}$ SUBATECH, IMT Atlantique, Nantes Universit\'{e}, CNRS-IN2P3, Nantes, France\\
$^{100}$ Sungkyunkwan University, Suwon City, Republic of Korea\\
$^{101}$ Suranaree University of Technology, Nakhon Ratchasima, Thailand\\
$^{102}$ Technical University of Ko\v{s}ice, Ko\v{s}ice, Slovak Republic\\
$^{103}$ The Henryk Niewodniczanski Institute of Nuclear Physics, Polish Academy of Sciences, Cracow, Poland\\
$^{104}$ The University of Texas at Austin, Austin, Texas, United States\\
$^{105}$ Universidad Aut\'{o}noma de Sinaloa, Culiac\'{a}n, Mexico\\
$^{106}$ Universidade de S\~{a}o Paulo (USP), S\~{a}o Paulo, Brazil\\
$^{107}$ Universidade Estadual de Campinas (UNICAMP), Campinas, Brazil\\
$^{108}$ Universidade Federal do ABC, Santo Andre, Brazil\\
$^{109}$ Universitatea Nationala de Stiinta si Tehnologie Politehnica Bucuresti, Bucharest, Romania\\
$^{110}$ University of Cape Town, Cape Town, South Africa\\
$^{111}$ University of Derby, Derby, United Kingdom\\
$^{112}$ University of Houston, Houston, Texas, United States\\
$^{113}$ University of Jyv\"{a}skyl\"{a}, Jyv\"{a}skyl\"{a}, Finland\\
$^{114}$ University of Kansas, Lawrence, Kansas, United States\\
$^{115}$ University of Liverpool, Liverpool, United Kingdom\\
$^{116}$ University of Science and Technology of China, Hefei, China\\
$^{117}$ University of Silesia in Katowice, Katowice, Poland\\
$^{118}$ University of South-Eastern Norway, Kongsberg, Norway\\
$^{119}$ University of Tennessee, Knoxville, Tennessee, United States\\
$^{120}$ University of the Witwatersrand, Johannesburg, South Africa\\
$^{121}$ University of Tokyo, Tokyo, Japan\\
$^{122}$ University of Tsukuba, Tsukuba, Japan\\
$^{123}$ Universit\"{a}t M\"{u}nster, Institut f\"{u}r Kernphysik, M\"{u}nster, Germany\\
$^{124}$ Universit\'{e} Clermont Auvergne, CNRS/IN2P3, LPC, Clermont-Ferrand, France\\
$^{125}$ Universit\'{e} de Lyon, CNRS/IN2P3, Institut de Physique des 2 Infinis de Lyon, Lyon, France\\
$^{126}$ Universit\'{e} de Strasbourg, CNRS, IPHC UMR 7178, F-67000 Strasbourg, France, Strasbourg, France\\
$^{127}$ Universit\'{e} Paris-Saclay, Centre d'Etudes de Saclay (CEA), IRFU, D\'{e}partment de Physique Nucl\'{e}aire (DPhN), Saclay, France\\
$^{128}$ Universit\'{e}  Paris-Saclay, CNRS/IN2P3, IJCLab, Orsay, France\\
$^{129}$ Universit\`{a} degli Studi di Foggia, Foggia, Italy\\
$^{130}$ Universit\`{a} del Piemonte Orientale, Vercelli, Italy\\
$^{131}$ Universit\`{a} di Brescia, Brescia, Italy\\
$^{132}$ Variable Energy Cyclotron Centre, Homi Bhabha National Institute, Kolkata, India\\
$^{133}$ Warsaw University of Technology, Warsaw, Poland\\
$^{134}$ Wayne State University, Detroit, Michigan, United States\\
$^{135}$ Yale University, New Haven, Connecticut, United States\\
$^{136}$ Yildiz Technical University, Istanbul, Turkey\\
$^{137}$ Yonsei University, Seoul, Republic of Korea\\
$^{138}$ Affiliated with an institute formerly covered by a cooperation agreement with CERN\\
$^{139}$ Affiliated with an international laboratory covered by a cooperation agreement with CERN.\\

\end{flushleft} 